\documentclass[10pt,preprint2,authoryear,longnamesfirst]{aastex}
\shorttitle{Scattering Toward B1849$+$005}
\shortauthors{Lazio}

\received{2003 May~8}
\revised{2004 June~3}
\accepted{2004 June~7}
\paperid{MS~58214}
\cpright{PD}{2003}

\newcommand{\src}{\protect\objectname[]{B1849$+$005}}
\newcommand{\psr}{\protect\objectname[PSR]{PSR~B1849$+$00}}
\newcommand{\aips}{\textsc{aips}}
\newcommand{\mjybm}{\mbox{mJy~beam${}^{-1}$}}

\begin{document}

\title{On the Enhanced Interstellar Scattering Toward \src}
\author{T.~Joseph~W.~Lazio}
\affil{Code~7213, Naval Research Laboratory, Washington, DC  20735-5351}
\email{Joseph.Lazio@nrl.navy.mil}

\begin{abstract}
This paper reports new Very Large Array (VLA) and Very Long Baseline
Array (VLBA) observations of the extragalactic source \src\ (Galactic
coordinates, $\ell = 33\fdg44$, $b = +0\fdg21$) at frequencies
between~0.33 and~15~GHz and the re-analysis of archival VLA
observations at~0.33, 1.5, and~4.9~GHz.  The structure of this source
is complex, confirming previous suggestions, but interstellar
scattering dominates the structure of the central component at least
to~15~GHz.  An analysis of the phase structure functions of the
interferometric visibilities shows the density fluctuations along this
line of sight to be anisotropic (axial ratio $= 1.3$) with a
frequency-independent position angle, and having an inner scale of
roughly a few hundred kilometers.  The anisotropies occur on length
scales of order $10^{15}\,\mathrm{cm}\,(D/5\,\mathrm{kpc})$, which
within the context of certain magnetohydrodynamic turbulence theories
indicates the length scale on which the kinetic and magnetic energy
densities are comparable.  A conservative upper limit on the velocity
of the scattering material is 1800~km~s${}^{-1}$, based on the lack of
changes in the shapes of the 0.33~GHz images.  In the 0.33~GHz field
of view, there are a number of other sources that might also be
heavily scattered, which would suggest that there are large changes in
the scattering strength on lines of sight separated by a degree or
less.  Both \src\ and \psr\ are highly scattered, and they are
separated by only 13\arcmin.  If the lines of sight are affected by
the same ``clump'' of scattering material, it must be at least 2.3~kpc
distant.  However, a detailed attempt to account for the scattering
observables toward these sources (angular broadening of the
extragalactic source, pulse broadening of the pulsar, and upper limits
on the angular broadening of the pulsar) does not produce a
self-consistent set of parameters for a clump that can reproduce all
three measured scattering observables.  A clump of H$\alpha$ emission,
possibly associated with the \ion{H}{2} region
\objectname[]{G33.418$-$0.004}, lies between these two lines of sight, but it seems
unable to account for all of the required excess scattering.
\end{abstract}


\section{Introduction}\label{sec:intro}

Density fluctuations on subparsec scales in the interstellar plasma
produce refractive index fluctuations.  In turn these refractive index
fluctuations scatter radio radiation propagating through the plasma.
The result is a rich set of observables including angular broadening
of compact sources, pulsar intensity scintillations, and low-frequency
variability of extragalactic sources \citep[for reviews see][]{r90,n92}.

The density fluctuations responsible for interstellar scattering
appear to be well parameterized by a power-law power spectrum.  In
many cases the spectral index of the power spectrum is near~11/3, the
value predicted by the Kolmogorov theory of turbulence for neutral,
incompressible unmagnetized fluids.  As the interstellar plasma is
compressible, ionized, and magnetized, it would appear to violate
these assumptions.  There has been recent progress in understanding
how a Kolmogorov-like spectrum could arise in a magnetized plasma,
though \citep[and references within]{lg03}. Combined with the large
Reynolds number of the interstellar plasma, this suggests that the
density fluctuations arise from a turbulent process.  A small number
of lines of sight toward pulsars are notable counter-examples, having
spectral indices larger than 4 \citep{hwg85,wc87,grl88,cfc93} or an
index approaching 4 \citep{sg90,mgrb90,wns94,mmrj95}.  An index larger
than 4 arises from a random distribution of density fluctuations,
while an index approaching 4 is interpreted as a modification of the
power-law spectrum by an inner scale indicative of a dissipative scale
for the turbulence.  Thus, determination of the spectral index of the
density fluctuation spectrum can provide clues about the mechanisms by
which the density fluctuations originate.

\cite{wns94} showed that not only can a density spectral index be
obtained, but that multi-frequency observations may probe the
organization and orientation of the density fluctuations responsible
for scattering.  They showed that the axial ratio and position angle
of \objectname[]{Cyg~X-3} are frequency dependent, a frequency
dependence that they interpreted in terms of an increasingly ordered
magnetic field on smaller and smaller scales.  \cite{tmr98} showed
subsequently that the axial ratio and orientation of the scattering
disk of \objectname[NGC]{NGC~6334B} also exhibited a frequency
dependence.  Thus, scattering observations may be a probe of the
magnetic field on subparsec scales.

Conversely, OH masers seen toward the Galactic center exhibit some of
the most extreme anisotropic images seen anywhere in the Galaxy
\citep{vfcd92,fdcv94}.  While it is not possible to measure a
frequency dependence for the scattering disks of the OH masers, the
scattering disk of \objectname[]{Sgr~A*} exhibits no change in its
axial ratio or orientation over nearly 1.5 decades in frequency
\citep{lbekrm85,jaunceyetal89,alberdietal93,y-zcwmr94,malp-td99,rrvt99}.
It is not yet clear what the different frequency behavior for the
scattering diameters of \objectname[]{Cyg~X-3} and
\objectname[NGC]{NGC~6334B} vs.\ \objectname[]{Sgr~A*} represents.
Possibilities might include a difference between the scattering media
in the Galactic center and disk or a much stronger magnetic field in
the Galactic center.

The source \src\ (Galactic coordinates, $\ell = 33\fdg44$, $b =
+0\fdg21$) is seen along the third most highly scattered line of sight
known, after the Galactic center \citep{lc98} and
\objectname[NGC]{NGC~6334B} \citep{tmr98}.  \cite{smbc86} conclude
that it is extragalactic because it is a compact, flat-spectrum
variable source.  Further evidence that \src\ is extragalactic is
found by comparing its \ion{H}{1} absorption spectrum to that of the
pulsar \psr, which is 13\arcmin\ from \src.  It must be more distant
than \psr\ and is probably at least 25~kpc distant
\citep{vgsg82,dkhv83,cfkw88,fw90}.  \psr\ is also the most heavily
scattered pulsar known, with a pulse broadening time
of~0.2~\emph{seconds} at~1.4~GHz.  Thus, comparison of the line of
sight to \src\ and \psr\ can probe changes in the scattering medium
transverse to the line of sight on scales not normally accessible.

This paper reports multi-frequency observations of \src, spanning
nearly two decades in frequency, that are used to reassess the structure
and nature of \src\ and constrain the scattering properties along this
line of sight.  In \S\ref{sec:observe} I describe the observations
and data reduction, in \S\ref{sec:scatter} I use these observations
to extract various scattering parameters along the line of sight to
\src, in \S\ref{sec:surround} I combine my observations with those from
the literature to probe the scattering environment around \src\ and
particularly toward \psr, and in \S\ref{sec:conclude} I summarize my
results.

\section{Observations and Analysis}\label{sec:observe}

The scattering disk of \src\ is large enough that it can be resolved
by a range of frequencies and baselines accessible with the VLA and
\hbox{VLBA}.  The details of the observations and data reduction for
the VLA and VLBA observations are sufficiently different that I
discuss them separately.  Table~\ref{tab:log} presents the observing
log.

\begin{deluxetable}{ccccccc}
\tablecaption{Observing Log\label{tab:log}}
\tabletypesize{\scriptsize}
\tablewidth{0pc}
\tablehead{
 \colhead{ } & \colhead{ } & \colhead{ } &
	\colhead{ } & \colhead{Recorded} & \colhead{On-Source} &
        \colhead{RMS Image} \\
 \colhead{Frequency} & \colhead{Epoch} & \colhead{Array} &
	\colhead{Bandwidth} & \colhead{Polarization} & \colhead{Time} &
	\colhead{Noise}\\
 \colhead{(GHz)} & \colhead{ } & \colhead{ } &
 \colhead{(MHz)} & \colhead{}  & \colhead{(hr)} & \colhead{(\mjybm)}}

\startdata
0.33 & 1987 August~20 & VLA (A-conf.) & 3.0 & R   & 0.28 & 80 \\
0.33 & 1998 March~19  & VLA (A-conf.) & 3.0 & R,L & 4.9  & 4.4 \\
1.5  & 1987 August~20 & VLA (A-conf.) & 43.75 & R & 0.375 & 0.32 \\
2.3  & 2002 August~23 & VLBA(FD, KP, OV, LA, PT) & 32 & R,L & 1.2 & $\ldots$ \\
4.9  & 1987 August~20 & VLA (A-conf.) & 50 & R  & 0.017 & 1 \\
5.0  & 1998 March~29  & VLBA$+$VLA1 & 56  & L   & 1.4 & 0.39 \\
8.4  & 1998 March~29  & VLBA$+$VLA1 & 64  & R   & 0.89 & 0.51 \\
15   & 1998 March~29  & VLBA$+$VLA1 & 64  & L   & 1.4 & 0.84 \\
22   & 1998 March~29  & VLBA$+$VLA1 & 64  & L   & 1.4 & $\ldots$ \\
\enddata
\end{deluxetable}

\subsection{VLA Observations (0.33--4.9~GHz)}\label{sec:vla}

\subsubsection{0.33~GHz}

Observations were obtained at~0.33~GHz with the A-configuration VLA on
1998 March~19.  The sources \objectname[3C]{3C48}
and~\objectname[3C]{3C286} were used in calibrating the visibility
amplitudes, and the VLA calibrator \objectname[VLA]{J1859$+$1259} was
used in calibrating the visibility phases.  Standard calibration and
radio frequency interference (RFI) excision procedures were used
within the NRAO Astronomical Image Processing System (\aips).  Several
iterations of self-calibration and imaging were used to approach the
thermal noise level.  During imaging, a polyhedron algorithm
\citep{cp92} was used to account for the non-coplanar nature of the
\hbox{VLA}.

At~0.33~GHz the field of view is large (2\fdg5 FWHM) and contains many
sources.  In order to isolate the visibilities from \src, the
following procedure was used.  Sources within the primary beam field
of view as well as strong sources outside the primary beam were
identified in the NRAO VLA Sky Survey \citep{ccgyptb98}.  Seventy
three such sources were identified, including \src.  Following
\cite{p89}, the Fourier transform of all of the sources was subtracted
from the visibility data, a modest amount of additional data editing
was performed, and the Fourier transform of only \src\ was added back
to the visibility data.

The imaging was done with the visibility data at the spectral
resolution at which it was acquired (97.7~kHz channels), both to avoid
bandwidth smearing and to make RFI easier to identify.  After removing
all other sources, bandwidth smearing is no longer an issue, and the
data were averaged over frequency.  Also, in order to reduce the data
volume before further processing, the visibilities were averaged
from~10~s to~240~s.  This time and frequency averaging also further
``dilutes'' the impact of any unsubtracted sources or errors in the
models of the sources that were subtracted.

I also obtained from the VLA Archive (program AS305) A-configuration VLA
observations of \src\ acquired originally on~1987 August~20 by
\cite{fsc91}.  The calibration of these observations was similar (but
with the VLA calibrator \objectname[VLA]{B1938$-$155} used to
calibrate the phases), but, given the relatively limited amount of
visibility data (15~min.\ observation), only \src\ was imaged, rather
than using targeted facetting.  The limited visibility data also
implies a higher noise level and limited image fidelity.

\begin{figure}
\epsscale{0.9}
\plottwo{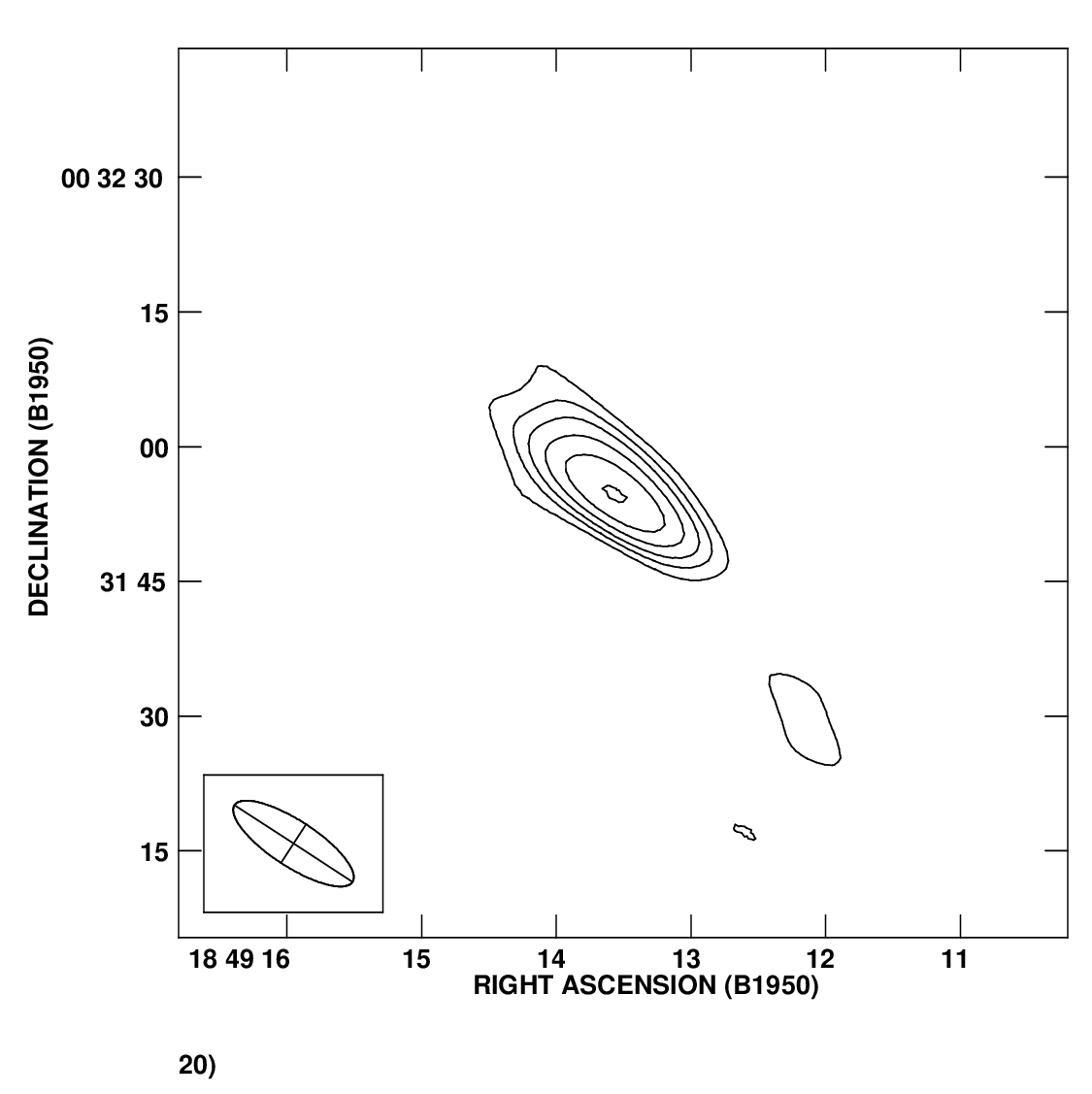}{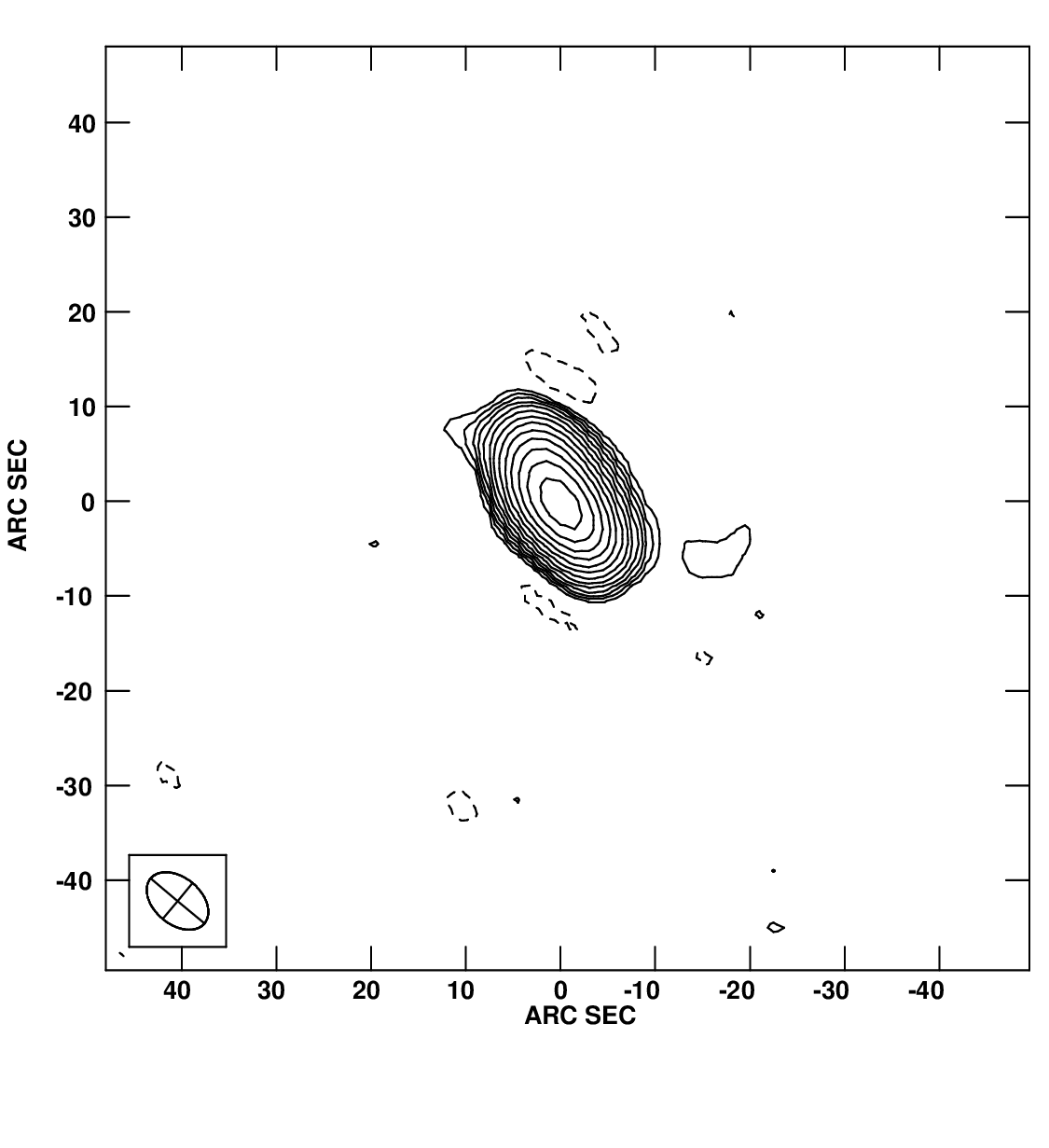}
\vspace{-0.5cm}
\caption[]{\src\ at~0.33~GHz.  
\textit{Left} The image resulting from the 1987 August~20
observations.  The rms noise level is 80~\mjybm, and the contour
levels are 80~\mjybm\ $\times$ $-3$, 3, 5, 7.07, 10, \ldots.  The
beam is $15\farcs7 \times 5\farcs1$ and is shown in the lower left.
\textit{Right} The image resulting from the 1998 March~19
observations.  The rms noise level is 4.4~\mjybm, and contour levels
are 4.4~\mjybm\ $\times$ $-3$, 3, 5, 7.07, 10, \ldots.  The beam is
$7\farcs4 \times 4\farcs9$ and is shown in the lower left.}
\label{fig:Pimg}
\end{figure}

Figure~\ref{fig:Pimg} shows the resulting 0.33~GHz images of \src.

\subsubsection{1.5~GHz}\label{sec:vla20}

The nominal resolution of the A-configuration VLA at 1.5~GHz is
approximately 1\farcs5.  The expected scattering diameter of \src\ at
this frequency is approximately 0\farcs3.  ``Super-resolution''
imaging can provide better than the nominal resolution by appropriate
weighting of the visibility data at the cost of a modest reduction in
sensitivity \citep{bss99}.  Given the strength of this source ($\sim
0.5$~Jy at~1.5~GHz), this is an acceptable trade-off.

I obtained from the VLA Archive (program AS305) A-configuration VLA
observations of \src\ acquired originally on~1987 August~20 by
\cite{fsc91} with the aim of imaging the source at super-resolution.
The source \objectname[3C]{3C~286} was used to calibrate the
visibility amplitude, and the VLA calibrator
\objectname[VLA]{B1821$+$107} was used in calibrating the visibility
phases.  Standard calibration procedures were used, including a number
of iterations of phase self-calibration.  

As was done in imaging the 0.33~GHz observations, a polyhedral
algorithm was used to account for the non-coplanarity of the VLA and
image the entire field of view (30\arcmin\ FWHM).  In practice, the
number of other sources in the field was sufficiently small and their
flux densities sufficiently low that ignoring the VLA's
non-coplanarity would have made little difference in the final image.

\begin{figure}
\epsscale{0.9}
\plotone{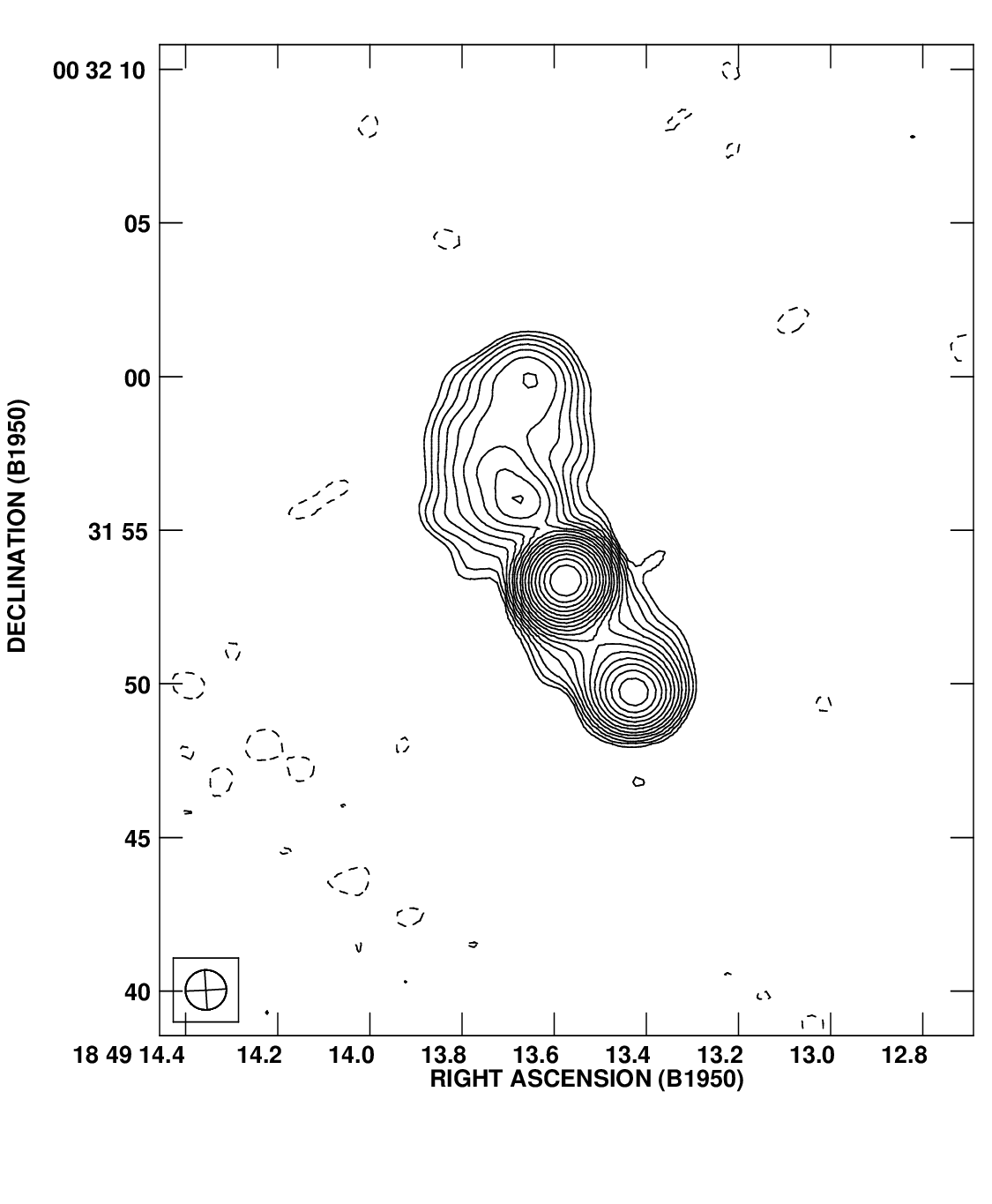}
\vspace{-1cm}
\caption[]{\src\ at~1.5~GHz, imaged using a modest amount of
super-resolution to obtain a 10\% reduction in the nominal beam
diameter.  The rms noise level is 0.32~\mjybm, and contour
levels are 0.32~\mjybm\ $\times$ $-3$, 5, 7.07, 10, 14.1, 20, \ldots.
The beam is $1\farcs3 \times 1\farcs3$ and is shown in the lower
left.}
\label{fig:Limg}
\end{figure}

Experiments with the amount of super-resolution used showed that
reasonable choices for the amount of weighting led to resolution
improvements by as much as 20\% (i.e., decrease in beam diameter).
Figure~\ref{fig:Limg} shows the resulting image of \src\ with a modest
amount of super-resolution (10\% reduction in the beam diameter).  In
practice, the amount of superresolution used made little difference in
the measured image diameter, but the measured image diameter is
smaller than the beam diameter.  Decreasing the beam diameter
increases the ratio of the image diameter to the beam diameter and
therefore increases the robustness of the measured image diameter,
particularly if the image diameter is smaller than the beam diameter,
as is the case here.

\subsubsection{4.9~GHz}\label{sec:vla6}

\begin{figure}
\epsscale{0.9}
\plotone{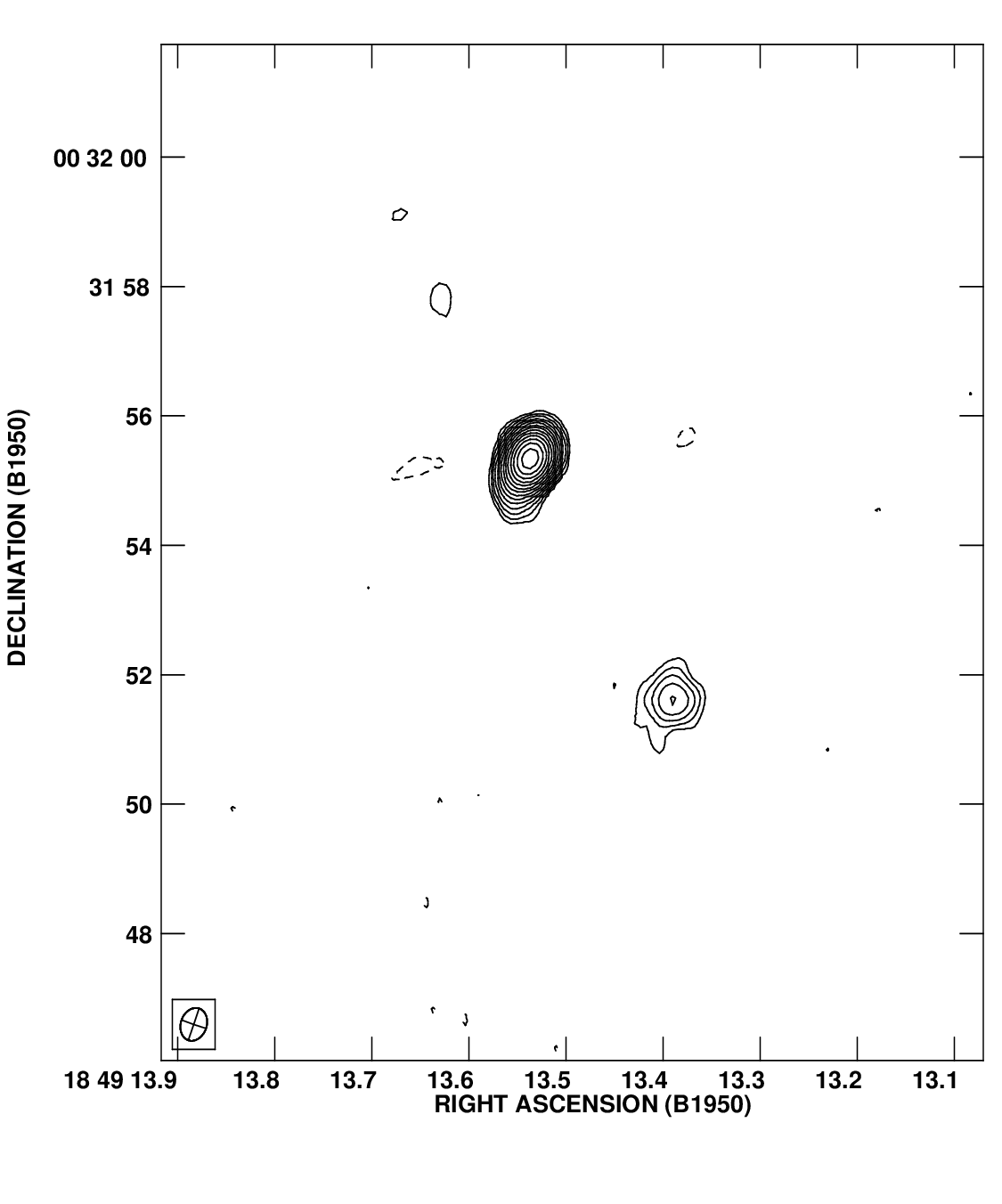}
\vspace{-0.75cm}
\caption[]{\src\ at~4.9~GHz, as observed by the \hbox{VLA}.  The rms
noise level is 1~\mjybm, and contour levels are 1~\mjybm\ $\times$
$-3$, 5, 7.07, 10, 14.1, \ldots.  The beam is $0\farcs5 \times 0\farcs4$ and is shown in the lower left.}
\label{fig:Cimg_VLA}
\end{figure}

\cite{fsc91} also obtained a single hour angle (1~min.\ duration) on
\src\ at~4.9~GHz during the same set of observations used to acquire
the 1.5~GHz observations.  Based on the results of imaging the 1.5~GHz
observations, these 4.9~GHz observations were re-analyzed also.  The
source \objectname[3C]{3C~286} was used to calibrate the visibility
amplitude, and the VLA calibrator \objectname[VLA]{B1821$+$107} was
used in calibrating the visibility phases.  Standard calibration
procedures were used, including a small number of iterations of phase
self-calibration.  Figure~\ref{fig:Cimg_VLA} shows the resulting
image.

\subsection{VLBA Observations (2.3--22~GHz)}\label{sec:vlba}

Observations at~5, 8.4, 15, and~22~GHz were obtained with the VLBA and
one antenna of the VLA on 1998 March~29.  Observations at~2.3~GHz were
obtained with only the inner antennas of the VLBA on 2002 August~23.
No fringes were detected at~22~GHz.  The lack of a detection at~22~GHz
is consistent with the lower sensitivity of the VLBA at this frequency
relative to the other frequencies observed.  At the other frequencies,
the correlated visibilities were fringe-fit and amplitude calibrated
in the standard fashion within \aips.  Fringe-fitting intervals were
2.5--3~min.  The calibrated visibilities were then exported to the
Caltech \texttt{difmap} program for hybrid mapping.

\begin{figure}
\epsscale{0.9}
\plotone{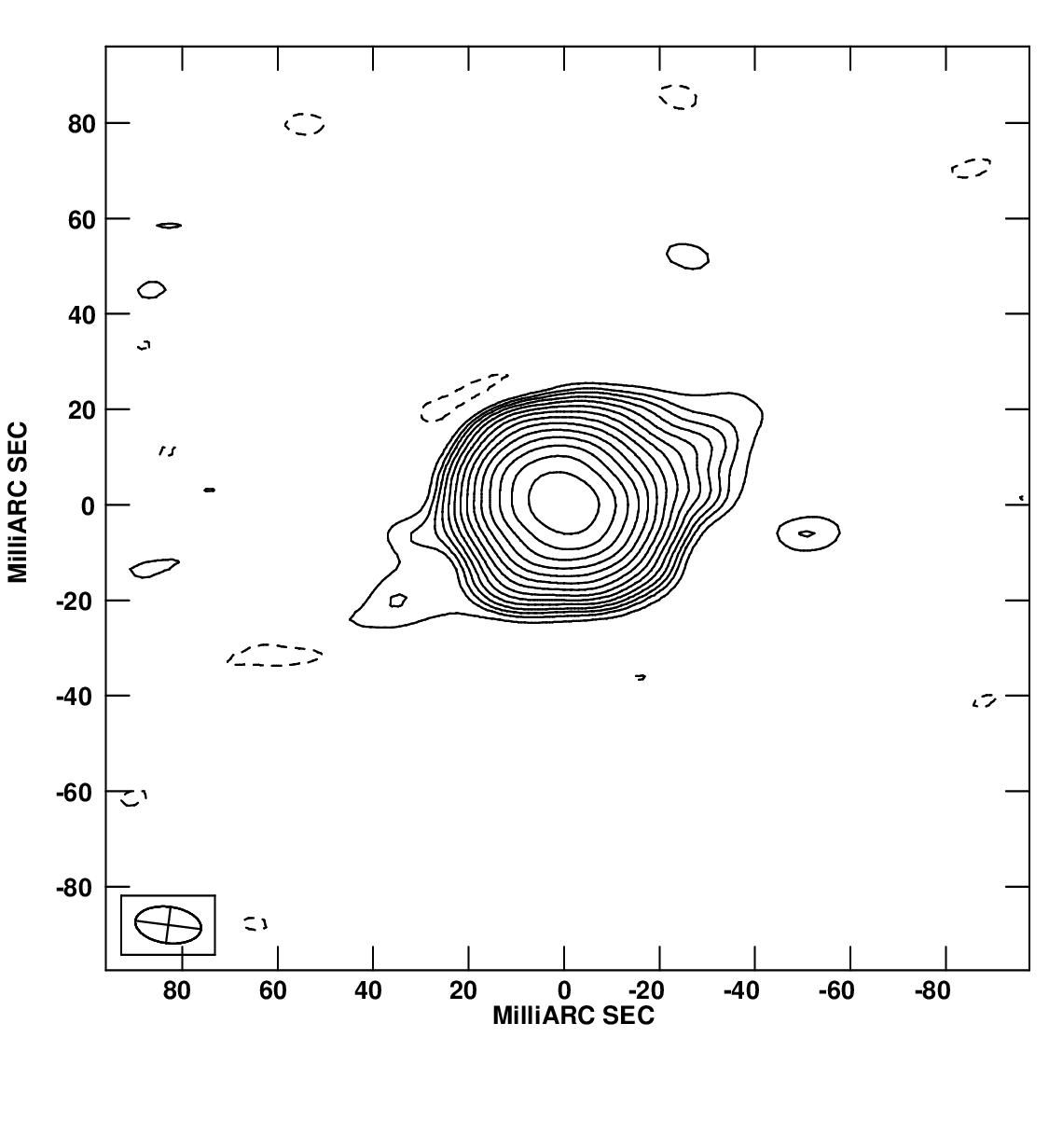}
\vspace{-1cm}
\caption[]{The central component of \src\ at~5~GHz, as observed by the
\hbox{VLBA}.  The rms noise level is 0.39~\mjybm, and contour levels
are 0.39~\mjybm\ $\times$ $-2$, 3, 5, 7.07, 10, \ldots.  The beam is
14~mas $\times$ 7.6~mas and is shown in the lower left.}
\label{fig:Cimg} 
\end{figure}

\begin{figure}
\epsscale{0.9}
\plotone{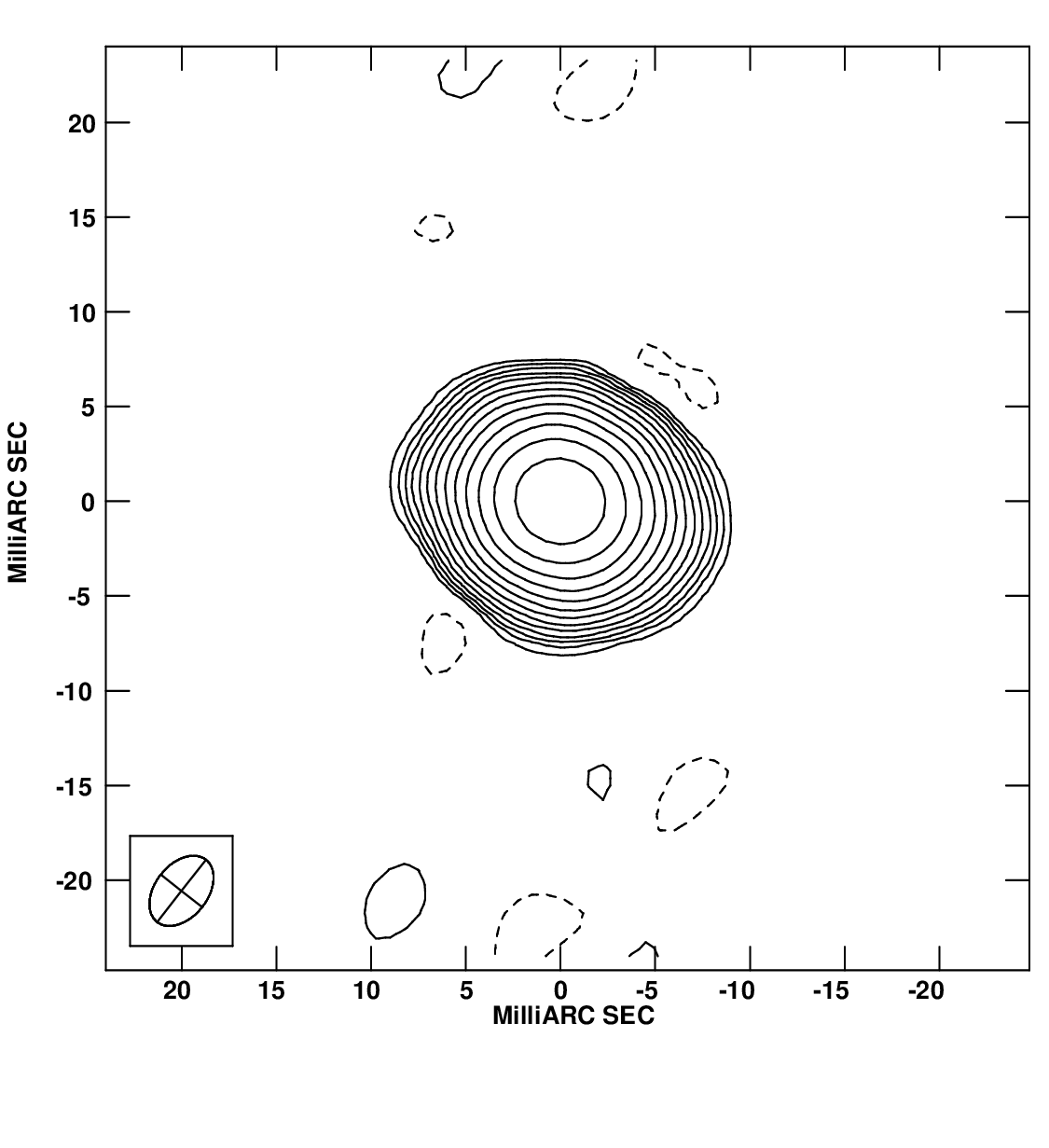}
\vspace{-1cm}
\caption[]{The central component of \src\ at~8.4~GHz.  The rms noise
level is 0.51~\mjybm, and contour levels are 0.51~\mjybm\ $\times$
$-3$, 3, 5, 7.07, 10, \ldots.  The beam is 4.2~mas $\times$ 2.8~mas
and is shown in the lower left.}
\label{fig:Ximg} 
\end{figure}

\begin{figure}
\epsscale{0.9}
\plotone{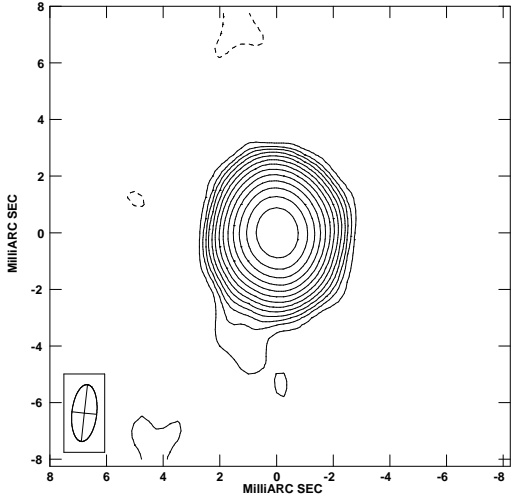}
\vspace{-1cm}
\caption[]{The central component of \src\ at~15~GHz.  The rms noise
level is 0.84~\mjybm, and contour levels are 0.84~\mjybm\ $\times$
$-2$, 3, 5, 7.07, 10, \ldots.  The beam is 2.0~mas $\times$ 0.88~mas
and is shown in the lower left.}
\label{fig:Uimg}
\end{figure}

For the 5, 8.4, and~15~GHz observations, after obtaining an image with
a noise near, but still well above, the thermal limit, I subtracted
the Fourier transform of \src\ from the visibility data, performed a
modest amount of additional data editing, and then added the Fourier
transform of \src\ back to the visibility data \citep{p89}.
Figures~\ref{fig:Cimg}--\ref{fig:Uimg} show the resulting images of
\src\ between~5 and~15~GHz.

At~2.3~GHz, a significant detection of \src\ was obtained only on the
shortest baseline, between the LA and~PT antennas.  As shown below,
this source has a fairly simple structure on milliarcsecond scales.
In order to estimate a diameter, the correlated visibility amplitude
on the LA-PT baseline was compared to total flux density of the
source.  \cite{frrr90} measure the flux density of the source
at~2.7~GHz to be 0.77~Jy; flux densities obtained by interpolating its
spectrum between~1.5 and~5~GHz (\citealt{fsc91}; this work) range
from~0.6 to~0.8~Jy.  Because the measurement by \cite{frrr90} utilized
a single-dish telescope, I adopt 0.77~Jy as its flux density at
2.3~GHz.  Assuming the source to be a circular Gaussian, its diameter
is then 110~mas (Table~\ref{tab:diameter}).

For further analysis, the data were averaged in time.  The averaging
times were 30~s at~2.3~GHz, 60~s at~5~GHz, 30~s at~8.4~GHz, and~15~s
at~15~GHz.

\section{Interstellar Scattering Toward \src}\label{sec:scatter}

\subsection{Relevant Scattering Formulae}\label{sec:formula}

The interferometric visibility of an infinitely distant point source
seen through a medium filled with random density fluctuations is
\begin{equation}
V(b) = e^{-D_\phi(b)/2},
\label{eqn:vis}
\end{equation}
for an interferometer with baseline length~$b$.  For the density power
spectrum relevant in the interstellar medium \citep{ars95}, the form
of the phase structure function depends upon the value of the
importance of~$l_1$ relative to the typical baseline length~$b$.
Anticipating later results, I consider two possibilities for
$D_\phi(b)$ \citep[e.g.,][]{cfrc87,cl91,lr99}
\begin{eqnarray}
& &D_\phi(b)= 8\pi r_e^2\lambda^2\,\mathrm{SM}f(\alpha) \times \nonumber\\
         && \quad\left\{
 \begin{array}{ll}
  b^{\alpha-2} & b \gg l_1,\\
  \Gamma(\alpha/2)\,l_1^{\alpha-2}[{}_1F_1(\alpha/2; 1; -(b/l_1)^2) - 1] & b \lesssim l_1.
 \end{array}
\right.
\label{eqn:dphi}
\end{eqnarray}
Here $r_e$ is the classical electron radius, $\lambda$ is the
free-space observing wavelength, $f(\alpha)$ is a slowly-varying
function of $\alpha$ with a value near unity, ${}_1F_1(a, b; x)$ is
the confluent hypergeometric function, and the scattering measure~SM
is proportional to the line-of-sight integrated rms electron density.
(Minor modifications are required if the source is not infinitely
distant, as assumed here.)  For reference, in the local interstellar
medium, $\mathrm{SM} \sim 10^{-3.5}$~kpc~m$^{-20/3}$ for a 1~kpc path
length.

The resulting scattering diameter, for a distant source and assuming
$\alpha = 11/3$ \citep{r90,cl91}, is
\begin{equation}
\theta_d = \left\{
 \begin{array}{ll}
 128\,\mathrm{mas}\ \mathrm{SM}^{0.6}\,\nu_{\mathrm{GHz}}^{-2.2} & b \gg l_1,\\
 71\,\mathrm{mas}\ \mathrm{SM}^{1/2}\,\nu_{\mathrm{GHz}}^{-2}\left(\frac{l_1}{100\,\mathrm{km}}\right)^{-1/6}                                        & b \ll l_1.
 \end{array}
\right.
\label{eqn:sdiam}
\end{equation}
where $\nu_{\mathrm{GHz}}$ is the observing frequency in GHz and SM is
in its canonical units of kpc~m${}^{-20/3}$.

\subsection{Angular Broadening and the Apparent Structure of \src}\label{sec:broaden}

On the basis of its spectrum, \cite{fsc91} suggested that the
structure of \src\ may be more complex than that of a single compact
component.  Figures~\ref{fig:Limg} and~\ref{fig:Cimg_VLA} demonstrate
that this is the case.  Although a strong central component
does dominate the structure of the source, there are weaker components
both to the southwest and northeast.  The components to the northeast
are at best only faintly visible in the VLA observations at~4.9~GHz
(Figure~\ref{fig:Cimg_VLA}), though their absence probably results
from the limited hour-angle coverage and surface brightness
sensitivity of the A-configuration VLA at this frequency.  The central
component may show a hint of a jet emerging to the southeast and
bending toward the southwest.  The existence of a jet must be
considered tentative, though, given the extremely limited hour angle
coverage (only 1~min.\ observation).  As the 0.33~GHz image is
elongated in approximately the same orientation as that of the central
and southwest components, I have also re-imaged the (1998 March~19)
0.33~GHz observations utilizing super-resolution.  There is no
indication of this second component, though the orientation of the
beam is similar to that of the central and southwest components.

\begin{figure}[tbh]
\epsscale{0.7}
\rotatebox{-90}{\plotone{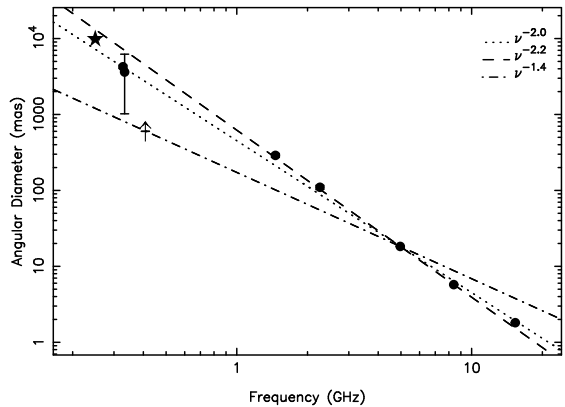}}
\caption[]{The angular diameter of \src\ as a function
of observing frequency.  The angular diameter was obtained by fitting
a Gaussian either to the image or to the visibility; if an elliptical
Gaussian was used in the fit, shown is the geometric mean of the major
and minor axes.  At most frequencies, the formal uncertainties are
comparable to the size of the points.  The datum at~0.25~GHz is from
P.~Rao~(2001, private communication) and that at~0.408 from
\cite{dbtbbc84}.  For reference, three curves are also plotted, one
showing a $\nu^{-2}$ dependence for the diameter (dotted line), one
showing a $\nu^{-2.2}$ dependence (dashed line), and one showing a
$\nu^{-1.4}$ dependence (dot-dash line) that would be expected if the
angular broadening for \src\ is anomalous in a manner similar to that
for \psr\ (\S\ref{sec:pulsar}).  All three curves are constrained to go
through the diameter measured at~5~GHz.}
\label{fig:diameter}
\end{figure}

The expressions of \S\ref{sec:formula} assume an infinitely distant
point source.  To what extent can they be applied to \src, given its
complex structure?  Figure~\ref{fig:diameter} shows the diameter of
the central component of \src\ as a function of observing frequency.
Although Figures~\ref{fig:Pimg}--\ref{fig:Uimg} show images of \src,
diameters measured from these images were not used in constructing
Figure~\ref{fig:diameter}.  Rather, Gaussian models were fit to the
visibility data, except in the case of the 2.3~GHz observations for
which a circular gaussian was assumed (\S\ref{sec:observe}).  Also
plotted in Figure~\ref{fig:diameter} are measurements at various other
frequencies.  At these other frequencies (0.25 and~0.408~GHz) the
visibility data were not available for analysis (as below), but
angular diameters could be determined.  For completeness,
Table~\ref{tab:diameter} summarizes the various fits for the diameter
of the source.

\begin{deluxetable}{lccccc}
\tablecaption{Gaussian Model Fits to the Central Component of \src\label{tab:diameter}}
\tabletypesize{\footnotesize}
\tablewidth{0pc} 
\tablehead{
 \colhead{Frequency} & \colhead{$S$} & \colhead{$a$} & \colhead{$b$} & \colhead{$\Phi$} & \colhead{Reference} \\
\colhead{(GHz)} & \colhead{(Jy)} & \colhead{(mas)} & \colhead{(mas)} & \colhead{(\arcdeg)}}
\startdata

 0.25 & 1.624 $\pm$ 0.009 & 11\,400 $\pm$ 70   &  8430   $\pm$ 60   & 44.2 $\pm$  0.7 & 2\\
 0.33 & 2.1   $\pm$ 0.2   &  6300   $\pm$ 1000 &  2080   $\pm$ 100  & 18   $\pm$  7   & 1\\
 0.33 & 2.12  $\pm$ 0.02  &  7050   $\pm$ 350  &  2560   $\pm$ 100  & 25.8 $\pm$  0.7 & 1\\
 0.41 & 1.7   $\pm$ 0.2   & $> 600$            &  \nodata           &  \nodata        & 3\\
 1.5\tablenotemark{a} & 0.63 $\pm$ 0.06 & 440 $\pm$ 6 &  190 $\pm$ 6  & 161  $\pm$  1   & 1\\
 2.3  & 0.77  $\pm$ 0.1   & 110     $\pm$ 5    &  110    $\pm$ 5    &  \nodata        & 1\\
 5.0  & 0.739 $\pm$ 0.009 & 19.34   $\pm$ 1    & 17.32   $\pm$ 1    & 27.0 $\pm$  0.6 & 1\\
 8.4  & 0.634 $\pm$ 0.006 &  6.71   $\pm$ 0.3  &  4.91   $\pm$ 0.2  &$-33.2 \pm$  0.3 & 1\\
15    & 0.831 $\pm$ 0.008 &  2.017  $\pm$ 0.1  &  1.623  $\pm$ 0.08 &$-21.1 \pm$  0.5 & 1\\

\enddata
\tablenotetext{a}{Only the central component was fit.}
\tablecomments{The visibility data are fit with a single Gaussian
model having a flux density~$S$, major axis~$a$, minor axis~$b$, and
position angle~$\Phi$.  In some cases a circular Gaussian was fit; no
position angle is reported for these cases.  The first entry
for~0.33~GHz is for the 1987 August~20 observations; the second is for
the 1998 March~19 observations.  The stated uncertainties are the
formal statistical uncertainties for the fits; the actual
uncertainties due to systematic effects are almost certainly larger.}
\tablerefs{(1)~This work; (2)~A.~P.~Rao~(2002, private communication); (3)~\cite{dbtbbc84}}
\end{deluxetable}

Figure~\ref{fig:diameter} also shows three curves, $\theta \propto
\nu^{-2.2}$, $\theta \propto \nu^{-2}$, and $\theta \propto
\nu^{-1.4}$.  The first two are relevant for strongly scattered
sources, with the first being appropriate if the electron density
spectrum is a power law with a Kolmogorov spectral index, $\alpha =
11/3$, while the second is appropriate if the range of observing
baselines is comparable to $l_1$.  The third curve shows the frequency
dependence for the angular broadening if scattering toward \src\ is
anomalous in a manner similar to that seen for the pulsar \psr\
(\S\ref{sec:pulsar}).  The relatively good agreement between the
angular diameters and the first two curves is an indication that
scattering dominates the apparent structure of the central component
at least up to~15~GHz.  Interpolating the angular diameter to the
fiducial frequency of~1~GHz, the scattering diameter of \src\ is
0\farcs6.  Based on their more limited set of observations,
\cite{fsc91} reached a similar conclusion: \src\ is heavily scattered,
though exhibits a more complex structure than simply a single compact
component.

In addition, the angular diameter of the southwest component was
measured, using the images shown in Figures~\ref{fig:Limg}
and~\ref{fig:Cimg_VLA}.  Its diameter is approximately constant
between~1.5 and~4.9~GHz, indicating that its diameter is much larger
than the scattering diameter.  This is consistent with its somewhat
diffuse appearance at~4.9~GHz.

\subsection{The Electron Density Spectrum}\label{sec:sf}

\begin{figure}[tbh]
\epsscale{0.7}
\rotatebox{-90}{\plotone{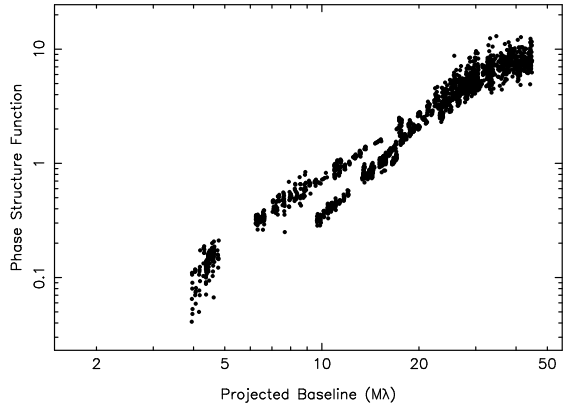}}
\caption[]{Structure function for the 8.4~GHz visibility data using
the isotropic baseline.  A small number of visibilities
below~2~M$\lambda$ are not shown.}
\label{fig:xsf}
\end{figure}

A more profitable analysis of the observations is to use the structure
function (equations~\ref{eqn:vis}--\ref{eqn:dphi}) to extract
parameters of the electron density spectrum.  My analysis follows
closely that of \cite{sc98}.  Figure~\ref{fig:xsf} shows an example of
a structure function constructed from the 8.4~GHz visibility data.

Clearly apparent in Figure~\ref{fig:xsf} is a non-random scatter in
the structure function, e.g., around 10~M$\lambda$.  Moreover,
Figures~\ref{fig:Pimg} and~\ref{fig:Cimg}--\ref{fig:Uimg} show that
the scattered image has (varying) degrees of anisotropy.  This
anistropy arises presumably because the density fluctuations
responsible for the scattering are themselves anisotropic.

The anisotropy of the image and visibility data (and of the underlying
density fluctuations) are described by an orientation~$\Psi$ and
aspect ratio~$\eta$.  In fitting the visibility data to
equation~(\ref{eqn:vis}), the baseline~$b$ is replaced by the
\emph{rotundate} baseline
\begin{equation}
b_r^2 = (u\cos\Psi + v\sin\Psi)^2 + \eta^2(-u\sin\Psi + v\cos\Psi)^2
\label{eqn:rotundate}
\end{equation}
where $u$ and~$v$ are the traditional interferometer baseline
coordinates measured in units of the observing wavelength.  The
rotundate baseline is simply the baseline measured in an anisotropic
coordinate system that is rotated relative to that of the
interferometer's coordinate system.  In this rotundate coordinate
system, $\Psi$ is measured counterclockwise from the $u$~axis, in
contrast to the position image in an image, which is measured
counterclockwise from north.  Thus, there is a 90\arcdeg\ offset from
an image position angle and~$\Psi$.

\begin{figure}[bth]
\epsscale{0.7}
\rotatebox{-90}{\plotone{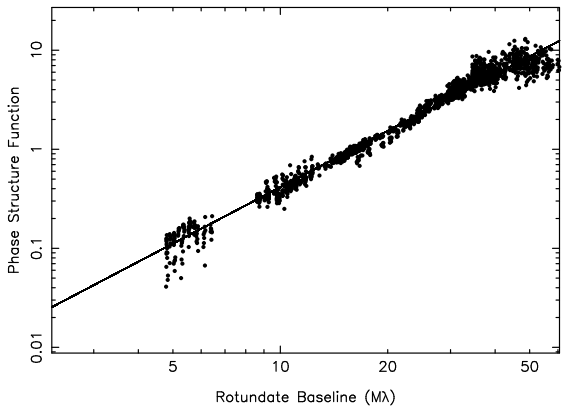}}
\caption[]{Structure function for the 8.4~GHz visibility data using
the rotundate baseline.  The solid line shows the best fitting model
for a power-law density spectrum (viz.\ Table~\ref{tab:fits}).  A
small number of visibilities below~2~M$\lambda$ are not shown.}
\label{fig:xsfrot}
\end{figure}

The non-random scatter in Figure~\ref{fig:xsf} arises because of the
use of the isotropic baseline ($\eta = 1$ and $\Psi = 0\arcdeg$).
Figure~\ref{fig:xsfrot} illustrates the use of the rotundate baseline
in plotting the structure function (using the best-fitting power-law
density spectrum model found below).  Clearly apparent is the dramatic
reduction in the scatter of the data; similar improvements are seen at
the other frequencies.  Figure~\ref{fig:xsfrot} also demonstrates that
the structure of \src\ does not have an adverse impact on efforts to
extract information about the scattering.  A single source component
does an excellent job of fitting the visibilities.  As might be
expected from Figures~\ref{fig:Limg} and~\ref{fig:Cimg_VLA}, the VLBI
observations are insensitive to the relatively low brightness
temperature components to the southwest and northeast.

One important aspect of the use of the rotundate baseline is that the
structure functions in equation~(\ref{eqn:dphi}) have been derived
assuming an isotropic density spectrum.  This \emph{ex post facto}
replacement of the rotundate baseline for the baseline violates that
assumption.  This replacement should be approximately valid provided
that $\eta$ is not much larger than unity.  Moreover, \cite{cb02} have
considered the propagation of radio waves through an anisotropic
medium.  They find that expressions like those of
equation~(\ref{eqn:dphi}) continue to hold; in particular, the
power-law dependences remain unchanged.

Using rotundate baselines and the structure functions of
equation~(\ref{eqn:dphi}), there are four or five parameters for which
to fit, depending upon the relative importance of the inner
scale---$\alpha$, SM, $\eta$, $\Psi$, and $l_1$.  I used a genetic
algorithm \citep{l97} to find the best fitting parameters in a
(unweighted) minimum chi-squared sense at each frequency.

I have fit the structure functions (as in Figure~\ref{fig:xsf}) for
the VLBI observations (2.3--15~GHz) and the 0.33~GHz VLA observations
of~1998 March~19.  The visibility data from the 1.5 or~4.9~GHz
observations from the VLA were not used because it is clear that the
source structure is complex and, at~4.9~GHz, the expected scattering
diameter is much smaller than what can be resolved by the \hbox{VLA}.
Also, the 0.33~GHz visibility data from the 1987 August~20 VLA
observations were not used because of their limited quantity and
restricted $u$-$v$ coverage.

Not all of the data were used in the fitting.  At long baselines, the
structure functions saturate while at short baselines they reach noise
plateaus (viz.\ equations~\ref{eqn:vis} and~\ref{eqn:dphi}).  The
saturation at long baselines is visible near~30~M$\lambda$ in
Figures~\ref{fig:xsf} and~\ref{fig:xsfrot}; the noise plateau at short
baselines occurs below~2~M$\lambda$ (for the 8.4~GHz data).
Consequently, limits were set at both short and long baselines.  After
excluding noisy data, the total number of data available to fit ranged
from several hundred to several thousand.

\begin{figure}
\epsscale{0.7}
\rotatebox{-90}{\plotone{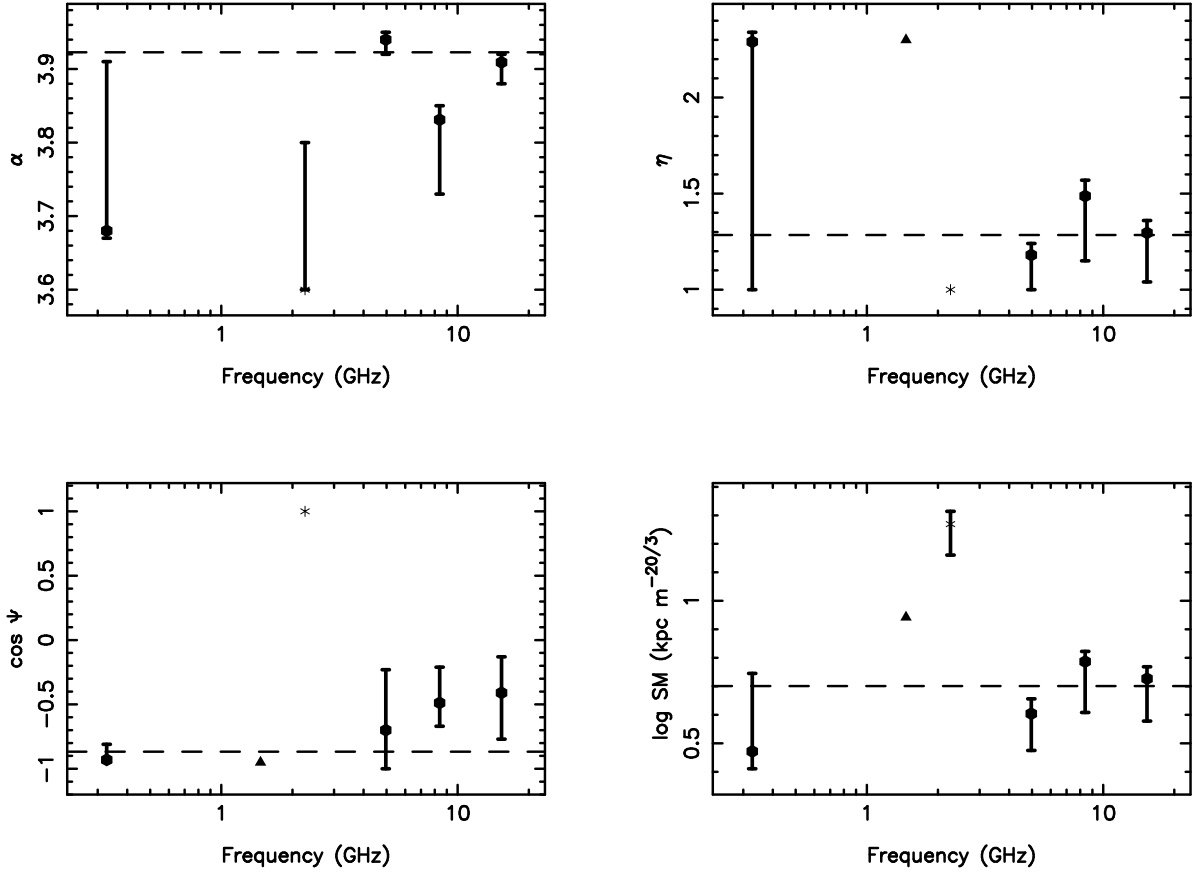}}
\caption[]{The best fit parameters as a function of frequency for a
power-law density spectrum.  The confidence regions shown are the
amount by which the parameter must change in order to cause the
chi-squared to increase by unity from its minimum value.  The
horizontal dotted line shows the weighted mean value for the
parameter.  Because \src\ was detected on only a single baseline
at~2.3~GHz, the anisotropy~$\eta$ and its orientation~$\Psi$ were
constrained to be 1 and~0 ($\cos\Psi = 1$), respectively.  At~1.5~GHz,
a structure function analysis was not conducted; the values shown are
derived from analysis of the image.}
\label{fig:fits}
\end{figure}

\begin{figure}
\epsscale{0.7}
\rotatebox{-90}{\plotone{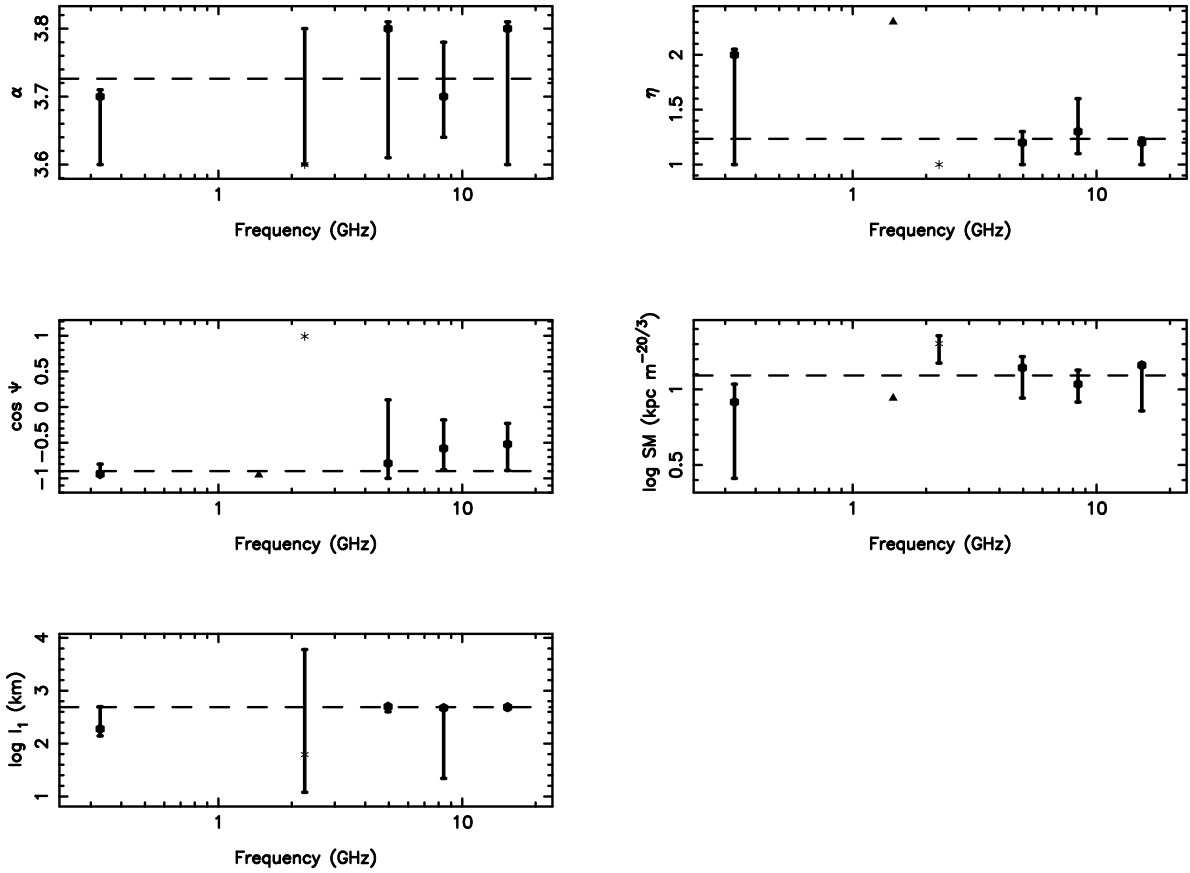}}
\caption[]{The best fit parameters as a function of frequency for a
density spectrum with an inner scale.  As for Figure~\ref{fig:fits}.}
\label{fig:fit1}
\end{figure}

Figures~\ref{fig:fits} and~\ref{fig:fit1} show the best fitting
parameters and their confidence ranges as a function of frequency for
both a power-law density spectrum ($b \gg l_1$) and a density spectrum
with an inner scale ($b \lesssim l_1$), respectively;
Tables~\ref{tab:fits} and~\ref{tab:fit1} tabulate the best-fitting
model parameters.  Figures~\ref{fig:fits} and~\ref{fig:fit1} also
indicate the (weighted) mean values, over all frequencies, for the
various parameters.  A $\chi^2$ statistic was used to test whether the
best fit-values are consistent with a single mean value, i.e., no
frequency dependence.

\onecolumn

\twocolumn

\begin{deluxetable}{ccccc}
\tablecaption{Power-law Density Spectrum Model Fitting\label{tab:fits}}
\tablewidth{0pc}
\tablehead{
 \colhead{Frequency} & \colhead{$\alpha$} & \colhead{$\eta$} &
	\colhead{$\Psi$} & \colhead{$\log\mathrm{SM}$} \\
\colhead{(GHz)} & \colhead{ } & \colhead{ } &
	\colhead{(\arcdeg)} & \colhead{(kpc~m${}^{-20/3}$)}}

\startdata
 0.33 & $3.68_{-0.01}^{+0.23}$ & $2.29_{-1.3}^{+0.05}$  & $158_{-14}^{+4}$  & $0.47_{-0.06}^{+0.41}$\\
 1.5\tablenotemark{a} & \nodata & 2.3                   & 161               & 0.98  \\
 2.3\tablenotemark{b} & 3.60   & \nodata                & \nodata           & 1.3 \\
 5.0  & $3.94_{-0.02}^{+0.01}$ & $1.18_{-0.18}^{+0.06}$ & $134_{-31}^{+46}$ & $0.60_{-0.15}^{+0.08}$\\
 8.4  & $3.83_{-0.10}^{+0.02}$ & $1.49_{-0.34}^{+0.08}$ & $119_{-17}^{+13}$ & $0.79_{-0.26}^{+0.08}$\\
15    & $3.91_{-0.03}^{+0.01}$ & $1.29_{-0.25}^{+0.07}$ & $114_{-17}^{+26}$ & $0.73_{-0.21}^{+0.07}$\\

\\
mean  & 3.92         & 1.30         & 141        & 0.68 \\
\enddata 
\tablenotetext{a}{At this frequency, the scattering parameters were
derived from the image, not from a structure function analysis.}
\tablenotetext{b}{At this frequency, the scattering diameter was
assumed to be circular.}
\end{deluxetable}

\begin{deluxetable}{cccccc}
\tablecaption{Power-law with Inner Scale Density Spectrum Model Fitting\label{tab:fit1}}
\tablewidth{0pc}
\tablehead{
 \colhead{Frequency} & \colhead{$\alpha$} & \colhead{$\eta$} &
	\colhead{$\Psi$} & \colhead{$\log\mathrm{SM}$} &
	\colhead{$l_1$} \\
\colhead{(GHz)} & \colhead{ } & \colhead{ } &
	\colhead{(\arcdeg)} & \colhead{(kpc~m${}^{-20/3}$)} &
	\colhead{(km)}}

\startdata
 0.33 & $3.70_{-0.10}^{+0.01}$ & $2.00_{-1.0}^{+0.05}$  & $160_{-17}^{+4}$  & $0.92_{-0.63}^{+0.28}$ & $190_{ -50}^{+310}$\\ 
 2.3  & 3.60                   & \nodata                & \nodata           & $1.30_{-0.1}^{+0.1}$   & $ 62_{ -12}$\\
 5.0  & $3.80_{-0.19}^{+0.01}$ & $1.20_{-0.20}^{+0.10}$ & $142_{-58}^{+38}$ & $1.14_{-0.42}^{+0.21}$ & $500_{-100}^{+0}$\\
 8.4  & $3.70_{-0.06}^{+0.08}$ & $1.30_{-0.20}^{+0.30}$ & $125_{-25}^{+26}$ & $1.03_{-0.25}^{+0.25}$ & $470_{-448}^{+30}$\\
15    & $3.80_{-0.20}^{+0.01}$ & $1.20_{-0.20}^{+0.04}$ & $121_{-18}^{+32}$ & $1.16_{-0.58}^{+0.04}$ & $490_{ -50}^{+10}$\\

\\
mean  & 3.78        & 1.23        & 148       & 1.14          & 470 \\
\enddata 
\tablenotetext{a}{At this frequency, the scattering diameter was
assumed to be circular.}
\end{deluxetable}

The confidence ranges for the parameters have been estimated both by a
local grid search and a non-linear least squares minimization around
the minimum $\chi^2$.  In the former method, the confidence range for
a parameter was determined by varying that parameter, while holding
all others fixed at their best fitting values, until the $\chi^2$
approximately doubled from its minimum value.  In the latter method a
Levenberg-Marquardt minimization routine was used to determine the
covariance matrix of the parameters.  The Levenberg-Marquardt
minimization yields confidence ranges much smaller than the grid
search.  We have adopted the confidence ranges for the grid search;
adopting smaller confidence ranges would make our following
conclusions more robust.

One aspect of Figures~\ref{fig:fits} and~\ref{fig:fit1} warrants
comment.  The various parameters---$\alpha$, SM, $\eta$, $\Psi$, and
$l_1$---are plotted as functions of frequency.  These parameters were
introduced as frequency-independent quantities
(equation~\ref{eqn:dphi}).  For instance, SM is proportional to the
rms electron density.  The apparent frequency dependence arises from
the use of interferometers whose baseline lengths were not scaled with
frequency.  For an interferometer with a typical (and fixed) baseline
length~$b$, it will resolve the scattering disk only if the scattering
disk is comparable in size to the typical angle measured by the
interferometer $\lambda/b$.  In turn if the scattering medium is at a
distance~$D$, the scattering disk will be produced by density
fluctuations on scales of $l^\prime \sim D(\lambda/b)$.  This length
scale is manifestly frequency dependent.  Thus, an apparent frequency
dependence in $\alpha$, SM, $\eta$, $\Psi$, or $l_1$ would reflect
different parameter values being obtained on different length scales
within the spectrum of scattering irregularities.  For brevity, in the
context of Figures~\ref{fig:fits} and~\ref{fig:fit1}, I shall continue
to speak of a frequency dependence for $\alpha$, SM, $\eta$, $\Psi$,
and $l_1$ with the understanding that any frequency dependence arises
from the different length scales probed in the scattering medium.

\subsubsection{Image Anisotropy}\label{sec:anistropy}

Regardless of whether a power-law or inner-scale density spectrum is
fit to the structure functions, the result confirms what the casual
inspection of Figures~\ref{fig:Pimg}--\ref{fig:Uimg} suggest: The
scattering toward \src\ is decidedly anisotropic.

At~0.33~GHz, the anisotropy appears large, $\eta \simeq 2.3$, but the
weaker components, especially the one to the southest, may be
contaminating the analysis.  The beam diameter at~0.33~GHz is
comparable to the separation between the components and the
orientation of the source at~1.5~GHz is similar to that of both the
source and the position angle of the beam at~0.33~GHz.  The
visibilities at~0.33~GHz appear consistent with only a single
component being present (\S\ref{sec:broaden} and
Figure~\ref{fig:diameter}), but the presence of the other components
cannot be excluded.  Indeed, convolving the 1.5~GHz image,
Figure~\ref{fig:Limg}, to a resolution comparable to that of the
0.33~GHz images, Figure~\ref{fig:Pimg}, produces an image similar to
the 0.33 GHz images.

Even if the analysis at~0.33~GHz is disregarded (and at~1.5
and~2.3~GHz, for which less comprehensive analyses could be
performed), the higher frequency data show a clear anisotropy.  The
fits at~5.0 and~15~GHz are just marginally consistent with an
isotropic image, while the fit at~8.4~GHz excludes an isotropic image.
The mean value for the image elongation is $\bar\eta = 1.2$.  This
value is comparable to that measured for \objectname[]{Cyg~X-3}
\citep[$\approx 1.3$,][]{wns94,mmrj95} and \objectname[NGC]{NGC~6334B}
\citep[1.2--1.5,][]{tmr98}, within the range of anisotropies 
measured for the extragalactic sources seen through the Cygnus region
\citep[1.1--1.8,][]{sc88,sc98,df01}, and somewhat smaller than the
value measured for the OH masers toward \objectname[W]{W49N}
\citep[2--3,][]{dgd94} and the Galactic center
\citep[$\approx 2.5$,][]{fdcv94}.  
\cite{rnb86} have predicted the anisotropy that would be induced, even if the
small-scale (diffractive) density fluctuations responsible for angular
broadening are isotropic, by refractive effects from large-scale
density fluctuations.  For a thin scattering screen with a Kolmogorov
density spectrum, the expected anisotropy is $\langle\eta\rangle <
1.05$.  The scattering toward \src\ is both strong and anisotropic,
and the anisotropy must result from the density fluctuations
responsible for the scattering.

Whether the position angle~$\Psi$ of the anistropy changes with
frequency, as \cite{wns94} and \cite{tmr98} found for
\objectname[]{Cyg~X-3} and \objectname[NGC]{NGC~6334B}, respectively,
is less certain.  In both Figure~\ref{fig:fits} and~\ref{fig:fit1}, an
apparent trend of a decreasing position angle (or increasing
$\cos\Psi$) is seen.  This is the case regardless of whether the
0.33~GHz datum is included or not.  However, as both figures show, the
uncertainties are also consistent with no change in position angle.

Using only the higher frequency data, the scattering diameter of \src\
ranges from~19~mas at~5.0~GHz to~2~mas at~15~GHz.  If the material
responsible for the scattering is at a distance of~5~kpc
(\S\ref{sec:pulsar}), the size of the region on which the anisotropic
density fluctuations occur must range from~$1.5 \times 10^{14}$--$1.4
\times 10^{15}$~cm (10--95~AU).  These scales are comparable to those
seen for \objectname[NGC]{NGC~6334B}, though this is due partially to
a selection effect in that the same telescope was used in both this
work and that of \cite{tmr98} so the range of accessible spatial
scales is the same.

Within the context of the models for magnetohydrodynamic turbulence
developed by Goldreich and collaborators \citep[e.g.,][]{lg01}, these
length scales
would represent the lower limit to their ``MHD scale.''  On this
length scale, the kinetic energy density ($\propto v^2$) of the
turbulent eddies has diminished to the point that the magnetic energy
density ($\propto B^2$) is comparable to it.  Thus, this length scale
marks where the turbulence makes a transition from isotropic on large
scale (comparable to the outer scale) to increasingly anisotropic on
small scales.

\subsubsection{Inner Scale}\label{sec:inner}

Comparison of Figure~\ref{fig:fits} and~\ref{fig:fit1} shows that the
density spectral index~$\alpha$ appears to change between~0.33~GHz
and~5.0~GHz for the power-law spectrum.  At~0.33~GHz, the fit for a
power-law spectrum finds $\alpha = 3.68$ while above~5.0~GHz, $\alpha >
3.8$.  In contrast, the fit for an inner scale spectrum reveals a much
more consistent set of values for~$\alpha$, with $3.7 < \alpha < 3.8$.
Moreover, the fits for the inner scale itself find consistent values
of a few hundred kilometers (though with large uncertainties).

I regard this as a detection of the effects of the inner scale on
scales of a few hundred kilometers.  The maximum baselines of the VLA,
used for the 0.33~GHz observation, are 35~km.  In contrast, the VLBA
baselines, used for the higher frequency observations, span the range
from a few hundred to a few thousand kilometers.  Thus the ``break''
between~0.33~GHz and~5.0~GHz in the fits for $\alpha$ in the power-law
density spectrum (Figure~\ref{fig:fits}) reflects the range of
baselines used in the observations.  

This value for the inner scale---a few hundred kilometers---is
comparable to that found toward other sources \citep{sg90,mgrb90,wns94,mmrj95}.

\subsection{Secular Change of the Scattered Image}\label{sec:vary}

If the plasma responsible for the scattering has a bulk (transverse)
velocity~$v$ (and it remains relatively unchanged as it moves across
the line of sight, ``frozen-flow'' approximation), changes in the
image appearance should appear on time scales comparable to the
refractive time scale $t_r = l_r/v$.  The two observations at~0.33~GHz
(Figure~\ref{fig:Pimg}) provide a 10.6~yr baseline over which to
search for secular changes in the image.

Table~\ref{tab:diameter} shows that both the fitted diameters and
position angles for the source do not change appreciably.  From both
sets of observations $\theta_d \simeq 4\arcsec$ at~0.33~GHz.  The
refractive scale is therefore $6 \times
10^{16}\,\mathrm{cm}\,D_{\mathrm{kpc}}$, where $D_{\mathrm{kpc}}$ is
the distance to the screen in kiloparsecs.  There is little change in
the image over the 10.6~yr interval.  The velocity of the scattering
material is therefore $v \le
1800\,\mathrm{km\,s}^{-1}\,D_{\mathrm{kpc}}$.  (Strictly, this
velocity is a combination of the velocities of the source, medium, and
observer.  Here the velocity of the source is assumed to be
effectively zero.)

This upper limit is conservative.  If the actual velocity were within
a factor of two, say, of this upper limit, some modest changes in at
least the orientation should start to become apparent.

The source \src\ is only 14\arcmin\ from the supernova remnant (SNR)
\objectname[]{G33.6$+$0.1}.  \cite{smbc86} observed \src\ with the aim
of assessing to what extent the SNR could contribute to the scattering
of \src.  While they inferred correctly that \src\ is heavily
scattered, they were reluctant to attribute all of the scattering to
\objectname[]{G33.6$+$0.1} given the low latitude line of sight to
\src.  Unfortunately, this upper limit on the velocity of the
scattering material sheds no light on whether material from the SNR is
responsible for the scattering.  The large scattering diameter
at~0.33~GHz means that relatively large length scales are probed
within the scattering material and that correspondingly long time
intervals are needed to place stringent constraints on the velocity.

The higher frequency VLBI observations may prove useful in
constraining the velocity of the scattering material in the future.
For instance, repeating the 8.4~GHz observations after a 7-yr interval
would place a 5~km~s${}^{-1}$ limit on the velocity (presuming that no
change in the scattering disk was seen).

\section{The Scattering Environment Around \src}\label{sec:surround}

\subsection{Nearby Lines of Sight}\label{sec:near}

By virtue of the VLA's large field of view at~0.33~GHz ($\approx
2\fdg5$), other highly scattered sources may be detectable in the
direction of \src.  With only a single frequency, I cannot demonstrate
unambiguously that scattering dominates the structure of other sources
in the field.  However, given that \src\ and \psr\ (see following
section) are both highly scattered and that
\objectname[PSR]{PSR~B1845$-$01} also exhibits a large amount of pulse
broadening \citep{rmdma97}, it is reasonable to consider that
scattering may contribute to the observed diameters of many of these
sources.

\begin{figure}[tbh]
\epsscale{0.9}
\rotatebox{-90}{\plotone{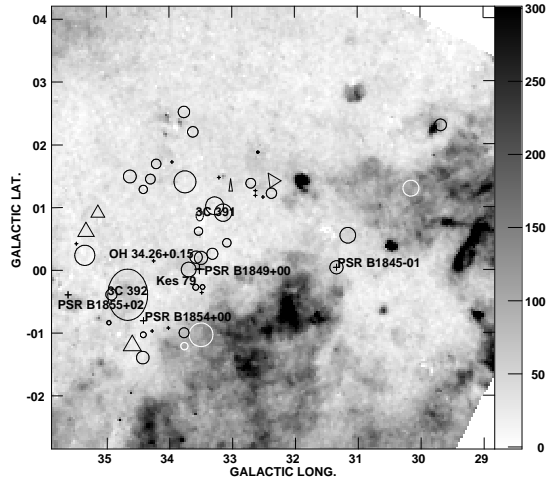}}
\caption[]{Lines of sight in the 0.33~GHz field of view of \src\ that
may also be dominated by scattering.  The gray scale shows the
smoothed, continuum-subtracted H$\alpha$ emission from the Southern
H$\alpha$ Sky Survey Atlas \citep{gmrv01} on a linear scale from~0
to~30~Raleighs.  Sources marked with crosses appear unresolved;
sources marked by circles appear resolved, with simple structure
indicating that scattering probably dominates; and sources marked by
triangles indicate sources whose structures may be more complex and
therefore contain contributions from both scattering and intrinsic
structure.  The size of a circle or triangle is proportional to the
source's diameter at~0.33~GHz.  Sources identified are the pulsars
\psr; \protect\objectname[PSR]{PSR~B1854$+$00};
\protect\objectname[PSR]{PSR~B1855$-$01}; the supernova remnants
\protect\objectname[Kes]{Kes~79}, \protect\objectname[3C]{3C~391}, and
\protect\objectname[3C]{3C~392}; and the maser
\protect\objectname[OH]{OH~34.26$+$0.15}.  The source \src\ is just
above \psr\ (viz.\ Figure~\ref{fig:closeup}).}
\label{fig:environment}
\end{figure}

Figure~\ref{fig:environment} shows the angular diameters for all
sources in the 0.33~GHz field of view of \src\ that are judged to be
compact and likely dominated by scattering.
Table~\ref{tab:environment} tabulates the diameters of these sources.

\begin{deluxetable}{cccccc}
\tablecaption{Other Possibly Scattered Sources Near
	\src\label{tab:environment}}
\tabletypesize{\small}
\tablewidth{0pc}
\tablehead{
  & &
\colhead{Galactic} & \colhead{Galactic} & & \\
\colhead{Right Ascension} & \colhead{Declination} & 
\colhead{Longitude} & \colhead{Latitude} & 
\colhead{Diameter} & \colhead{Structure} \\
\multicolumn{2}{c}{(J2000)} & & & \colhead{(\arcsec)}}

\startdata
18 35 10.617  &  01 45 52.86  &   32.650  &    4.422  &  $\ldots$  &  ? \\
18 37 18.258  & $-01$ 50 26.68 &  29.680  &    2.302  &   3.8      &    \\
18 41 44.789  & $-01$ 52 52.35 &  30.152  &    1.297  &   5.1      &    \\
18 43 58.583  &  01 53 27.68  &   33.766  &    2.522  &   3.7      &    \\
18 44 07.675  &  00 33 17.10  &   32.592  &    1.879  &   0.6      &    \\
\\	       
18 44 50.864  &  01 37 15.89  &   33.624  &    2.205  &   3.4      &    \\
18 45 17.789  &  00 07 42.94  &   32.117  &    1.308  &   5.2      &  ? \\
18 46 03.708  &  00 03 39.24  &   32.265  &    1.168  &   3.3      &    \\
18 46 05.681  &  00 25 38.53  &   32.703  &    1.384  &   3.1      &    \\
18 46 14.538  & $-01$ 19 40.83 &  31.158  &    0.550  &   5.0      &    \\
\\	       
18 46 23.171  &  00 18 28.74  &   32.630  &    1.264  &  $\ldots$  &    \\
18 46 31.286  &  00 09 08.35  &   32.507  &    1.163  &  $\ldots$  &    \\
18 46 39.510  &  00 16 23.36  &   32.630  &    1.188  &  $\ldots$  &    \\
18 46 41.266  &  00 55 14.98  &   31.571  &    0.637  &  $\ldots$  &    \\
18 46 53.543  &  00 41 02.80  &   33.023  &    1.323  &   2.6      &  ? \\
\\	       
18 47 09.999  &  01 41 55.15  &   33.957  &    1.724  &  $\ldots$  &    \\
18 47 43.843  &  01 54 33.63  &   34.209  &    1.695  &   3.0      &    \\
18 47 55.076  &  01 22 19.10  &   33.752  &    1.409  &   7.0      &    \\
18 48 23.488  & $-01$ 23 58.55 &  31.339  &    0.039  &   4.2      &    \\
18 48 25.281  &  00 46 36.09  &   31.897  &    0.317  &   5.7      &    \\
\\	       
18 48 34.187  &  00 36 18.25  &   32.067  &    0.362  &   5.5      &    \\
18 48 45.681  &  01 52 55.68  &   34.302  &    1.453  &   3.1      &    \\
18 49 12.392  &  02 11 32.05  &   34.629  &    1.496  &   4.1      &    \\
18 49 33.461  &  01 54 29.11  &   34.416  &    1.288  &   2.6      &    \\
18 50 10.398  &  00 20 04.02  &   33.086  &    0.434  &   2.7      &    \\
\\	       
18 50 20.140  &  00 49 13.91  &   33.537  &    0.619  &   2.6      &    \\
18 50 40.088  &  03 57 33.61  &   36.370  &    1.975  &   2.4      &    \\
18 51 52.819  &  00 40 01.42  &   32.389  &  $-0.403$ &   3.9      &    \\
18 52 12.978  &  02 22 52.19  &   35.140  &    0.912  &   5.1      &  ? \\
18 52 27.364  &  00 32 00.70  &   33.523  &    0.017  &  $\ldots$  &    \\
\\	       
18 53 23.434  &  00 21 37.42  &   32.834  &  $-0.599$ &   1.5      &    \\
18 53 36.039  &  00 27 14.06  &   32.775  &  $-0.688$ &   1.9      &    \\
18 53 37.212  &  02 24 56.68  &   35.331  &    0.616  &   6.0      &  ? \\
18 53 43.689  &  00 19 56.05  &   33.489  &  $-0.358$ &  $\ldots$  &    \\
18 54 34.997  &  02 27 51.16  &   35.484  &    0.424  &  $\ldots$  &    \\
\\	       
18 54 59.951  &  02 15 39.94  &   35.350  &    0.239  &   6.4      &    \\
18 56 09.016  &  00 02 04.43  &   33.500  &  $-1.033$ &   7.5      &    \\
18 56 27.229  &  01 35 57.48  &   34.927  &  $-0.386$ &   3.6      &    \\
18 56 31.711  &  00 17 27.47  &   33.771  &  $-1.000$ &   3.2      &    \\
18 56 43.059  &  00 32 50.93  &   34.021  &  $-0.925$ &  $\ldots$  &    \\
\\	       
18 57 16.674  &  00 11 21.39  &   33.766  &  $-1.213$ &   2.3      &    \\
18 57 21.043  &  00 45 28.03  &   34.281  &  $-0.970$ &  $\ldots$  &    \\
18 57 49.365  &  00 51 17.91  &   34.421  &  $-1.031$ &   1.8      &    \\
18 58 07.936  &  01 25 49.86  &   34.969  &  $-0.837$ &   1.3      &    \\
18 58 44.966  &  00 55 59.52  &   34.596  &  $-1.201$ &   6.6      &  ? \\
\\	       
18 59 08.028  &  00 41 48.99  &   34.430  &  $-1.394$ &   4.0      &    \\
18 51 12.962  &  00 27 37.89  &   33.317  &    0.259  &   3.5      &    \\
\enddata
\tablecomments{Column~5, Diameter, lists unresolved sources as $\ldots$.
Column~6, Structure, indicates whether a visual 
inspection of the source image suggests a contribution from intrinsic
structure to the measured angular diameter.}
\end{deluxetable}

If the structures of these sources are dominated by scattering, there
is substantial variation in the scattering on sub-degree angular
scales.  \cite{slyshetal01} have reached a similar conclusion by
comparing their space VLBI observations of the maser
\objectname[]{OH~34.26$+$0.15} to scattering measurements for \psr\
and \src.  They find the maser to have an angular diameter of less
than 1~mas, from which they deduce a scattering measure of~$1.68
\times 10^{-3}$~kpc~m${}^{-20/3}$, far less than that of \psr\ or
\src, though in this case it is likely that the maser lies in front of
the strong scattering region.

Figure~\ref{fig:environment} suggests that the region of intense, yet
spatially-variable scattering may subtend a large angle on the sky.
For instance, there are a number of sources near
(18${}^{\mathrm{h}}$~47\fm5, $+2\arcdeg$) whose diameters range from
unresolved to 7\arcsec.  Numerous other cases of unresolved sources
fairly close to sources with measurable diameters also can be
identified.  These lines of sight may be similar to other directions
showing both heavy and anisotropic scattering and large changes in the
scattering strength on degree-size scales such as the Cygnus region
\citep{fsm89,lsc90,sc98,df01}, the Galactic center
\citep{vfcd92,fdcv94}, and along the Galactic plane \cite{dbtbbc84,fsc91}.

\subsection{The Line of Sight to \psr}\label{sec:pulsar}

The pulsar \psr\ is only 13\arcmin\ from \src\ and is the most heavily
scattered pulsar known, with a pulse broadening time
of~0.2~\emph{seconds} at~1.4~GHz.  \cite{lkmll01} show that
between~1.4 and~2.7~GHz the pulse broadening of this pulsar is
anomalous in the sense that its frequency dependence is $\nu^{-x}$
with $x = 2.8^{+1.0}_{-0.6}$ rather than the expected $x \approx 4$.
In turn, this anomalous frequency scaling implies a density spectral
index of $\alpha = 6.6^{+7.8}_{-2.3}$.  (For a review of other
observations of this pulsar, see \citealt*{swdddha03}.)

My observations of \src\ are at frequencies that bracket those used by
\cite{lkmll01} to observe \psr\ (Figure~\ref{fig:diameter}).
Unfortunately the large uncertainties on the angular diameters
between~0.33 and~2.3~GHz preclude definite conclusions about the
possibility of similar anomalous scattering along the line of sight
to \src.  On the one hand, the frequency scaling for the angular
diameter between~1.5 and~2.3~GHz is $\nu^{-2.3}$ and the frequency
scaling between~0.33~GHz and~5~GHz is $\nu^{-2}$.  Both of these
frequency scalings (as well as Figure~\ref{fig:diameter}) would
suggest a frequency dependence of approximately $\nu^{-2}$ is a much
better description for the angular broadening than a dependence of
approximately $\nu^{-1.4}$ (because the pulse broadening and angular
broadening are related as $\tau \sim \theta^2$).  Recall, though, that
the 0.33~GHz diameter may be overestimated due to contamination from
the source structure (\S\ref{sec:broaden}), the 1.5~GHz diameter is
somewhat smaller than the resolution of the array, and at~2.3~GHz the
diameter has been estimated from only a single baseline.

\citet[][their Figures~2 and~3]{cl01} showed that anomalous pulse
broadening, over a limited frequency range and possibly consistent
with that observed by \cite{lkmll01}, can be produced by two
scattering screens having different scattering strengths.  Such a
scenario would produce anomalous angular broadening as well.  I model
this anomalous scattering in a piece-wise linear fashion, with angular
broadening scaling approximately as $\nu^{-2.2}$ below~1.4~GHz and
above~2.7~GHz while between~1.4 and~2.7~GHz it would scale as
$\nu^{-1.4}$.  This model produces an overall frequency scaling
between~0.33 and~5~GHz of $\nu^{-1.9}$.  Thus, while
Figure~\ref{fig:diameter} shows that the angular diameter appears to
scale as expected with frequency, a modest amount of anomalous
scattering at frequencies near~1.5~GHz cannot be excluded.

A more stringent constraint on the anomalous scattering of \psr\ could
be obtained from multi-frequency observations of its angular diameter.
The field of view of the VLA is sufficiently large that \psr\ is
detected in both the 0.33 and~1.5~GHz observations.  Unfortunately,
at~0.33~GHz it is unresolved, while at~1.5~GHz it is sufficiently near
the edge of the primary beam that it is difficult to obtain a reliable
angular diameter estimate.  Very long baseline interferometry
observations are under way (and will be reported elsewhere) to observe
\src\ and \psr\ simultaneously near~1.5~GHz in an effort to see if the
pulsar's angular diameter appears anomalous.

Nonetheless, by combining the various observations for \src\ and \psr,
it may be possible to constrain the properties of the regions
responsible for the excess scattering.  The scattering toward \src\
and \psr\ is a factor of~5 or more in excess of what is predicted by
the large-scale components---a thick disk, thin disk, and spiral
arms---of the Cordes-Lazio model \citep[and the model by][]{tc93}.  In
order to reproduce the observed scattering toward \psr, \cite{cl03}
introduced a ``clump'' or ``cloud'' of electrons along the line of
sight to it.  They require the clump to have an internal electron
density $n_e = 10$~cm${}^{-3}$ and a fluctuation parameter $F = 200$.
The fluctuation parameter~$F$ \citep[see also][]{cwfsr91} is a
combination of the fractional variation of the electron density internal
to the clump and the outer scale.  Larger values of~$F$ mean that a
clump is more efficient at scattering.  (For a clump, the fluctuation
parameter and the scattering measure are proportional, $\mathrm{SM}
\propto Fn_e^2\Delta r$, where $\Delta r$ is the size of the clump.)
They placed the clump at a distance of~7.4~kpc and assumed its size to
be $\Delta r = 20$~pc.

The power of combining the measurements is twofold.  First, the
magnitude of angular broadening is strongly dependent upon the
distribution of scattering material along the line of sight, with
material close to the observer contributing more strongly
\citep{vfcd92,lc98}.  Consider a Galactic source at a distance~$D$
with angular broadening~$\theta_{\mathrm{Gal}}$ and an extragalactic
source along a similar line of sight with angular
broadening~$\theta_{\mathrm{xgal}}$.  If the scattering along the line
of sight is dominated by a single screen at a (radial)
distance~$\Delta$ from the Galactic source, then \citep{vfcd92}
\begin{equation}
\frac{\theta_{\mathrm{Gal}}}{\theta_{\mathrm{xgal}}}
 = \frac{\Delta}{D}.
\label{eqn:scattering}
\end{equation}
This relationship implies that, even though seen through the same
scattering screen, a Galactic source will have a smaller scattering
diameter than an extragalactic source; this result occurs because the
scattering screen is less effective at scattering the diverging
wavefronts from the Galactic source than the planar wavefronts from
the extragalactic source.

The 0.33~GHz scattering diameter of \src\ is 4\farcs2.  Taking an
upper limit on the scattering diameter of \psr\ to be 3\arcsec\ ($=
1/2$ the beam diameter), $\Delta/D < 0.7$.  As \psr\ is at a distance
of roughly 8~kpc, $\Delta < 5.7$~kpc or that the scattering screen
must be more than 2.3~kpc from the Sun.

The second reason for combining the measurements toward \src\ and
\psr\ is that angular broadening and pulse broadening have different
weightings for the distribution of scattering material along the line
of sight.  If $x$ is the fractional distance between the pulsar and
the observer (with $x = 0$ being at the pulsar and $x = 1$ being at
the observer), then pulse broadening is weighted as $x(1-x)$ while
angular broadening is weighted as $x^2$.

I have repeated the analysis of \cite{cl03} but now requiring that a
single clump describe the excess scattering toward both \psr\ and
\src.  Specifically, I adopt the large-scale components of the
Cordes-Lazio model and then search for a clump of excess scattering
that can explain the observed scattering diameter for \src, the upper
limit on the scattering diameter of \psr, and the pulse broadening of
\psr.  A grid search over the three parameters $n_e$, $F$, and~$x$ was
used to search for the best fit in a minimum $\chi^2$ sense.  The size
of the clump is assumed to be 40~pc, so that it covers both lines of
sight even when the clump is near the pulsar.

An added constraint in the fitting process is that the dispersion
measure of the pulsar ($\mathrm{DM} = 680$~pc~cm${}^{-3}$) and the
modeled distance to the pulsar must be self-consistent.  That is,
increasing the density within the clump necessarily reduces the
modeled distance to the pulsar, which in turn affects the weighting
for the scattering observables.  The fitting procedure was therefore
the following.  The dispersion measure contributed by the clump,
$\delta\mathrm{DM}$, was determined for each set of values for~$n_e$
and $x$.  The model distance for the pulsar was determined by
requiring that the large scale components of the Cordes-Lazio model
account for the remainder of the pulsar's DM, i.e., for $(\mathrm{DM}
- \delta\mathrm{DM}$).  This new distance was then used in determining
the weighting factors for the modeled scattering observables.  In
practice, the $\delta\mathrm{DM}$ contributed by the clump is
sufficiently small that it does not affect the modeled pulsar distance
($\approx 9$~kpc) significantly.

No set of parameters can reproduce all three observational
constraints.  In particular, the pulsar's pulse broadening is
consistently underestimated (by about an order of magnitude) while the
extragalactic source's angular diameter is overestimated (by about a
factor of~2--3).  As might be expected, because the observational
constraints are all due to scattering, the parameters $n_e$ and~$F$
are highly correlated.  The best fit value for the distance to the
clump is 4.7~kpc.  While consistent with the crude estimate obtained
above based on just the scattering diameters, this distance for the
clump corresponds to $x \approx 1/2$.  This value simply reflects the
value of~$x$ that maximizes the modeled pulse broadening, which still
only poorly reproduces the observed pulse broadening.

If the excess scattering toward both \src\ and \psr\ is attributed to
the same clump, the angular diameters provide the most useful
constraint; the clump must be at least 2.3~kpc distant.  Given that no
single clump reproduces all of the observations, it is quite likely
that different clumps affect the two lines of sight.

\begin{figure}[hbt]
\epsscale{0.95}
\rotatebox{-90}{\plotone{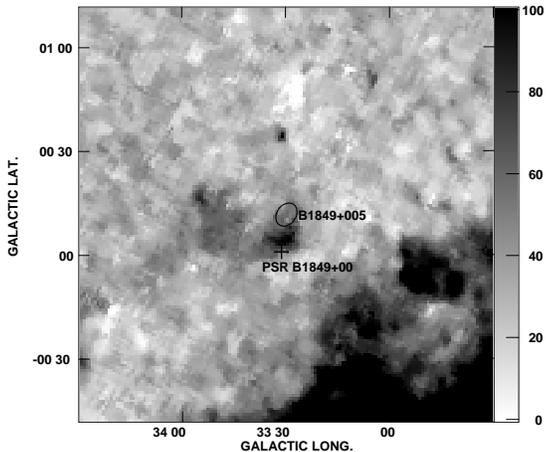}}
\vspace{-1cm}
\caption[]{A subimage of Figure~\ref{fig:environment} showing the
lines of sight to \src\ and \psr.  The gray scale shows the smoothed,
continuum-subtracted H$\alpha$ emission on a linear scale from~0
to~10~Raleighs.  The sizes of the symbols are proportional to the
scattering diameter (\src) or upper limit (\psr) at~1~GHz.  For the
scattering diameter of \src, it is shown with the correct orientation
and anisotropy.}
\label{fig:closeup}
\end{figure}

Is there a structure along the line of sight that could contribute to
anomalous scattering toward \psr\ but not \src?
Figure~\ref{fig:closeup} shows the H$\alpha$ emission in a small
region around these two sources taken from the Southern H$\alpha$ Sky
Survey Atlas \citep[SHASSA,][]{gmrv01}.  Clearly apparent between the
two, but offset slightly toward \psr, is a clump of slightly stronger
H$\alpha$ emission.

This H$\alpha$ clump appears to lie on the outskirts of the \ion{H}{2}
region \objectname[]{G33.418$-$0.004} \citep{l89}.  If this H$\alpha$
clump is associated with the \ion{H}{2} region, the H$\alpha$ clump
would represent only a small fraction of the total H$\alpha$
luminosity of the \ion{H}{2} region.  Based on the recombination line
strengths for the \ion{H}{2} region, much of it must be heavily
absorbed at optical wavelengths.  This line of sight is toward the
Aquila Rift \citep{dht01}, and the area to the south of the H$\alpha$
clump appears consistent with being heavily obscured
(Figure~\ref{fig:closeup}).

The emission measure of this \ion{H}{2} region is $\mathrm{EM} \approx
7000$~pc~cm${}^{-6}$ \citep{l89}.  The H$\alpha$ clump and the line of
sight to \psr\ appear to be where the 5~GHz continuum has dropped by
approximately a factor of~2 \citep{adps79}.  Thus, the EM contributed
by the \ion{H}{2} region along the line of sight to the pulsar is
perhaps 3500~pc~cm${}^{-6}$.  The equivalent scattering measure is
$\log\,\mathrm{SM} = 0.81$ [kpc~m${}^{-20/3}$], assuming an outer
scale of~1~pc \citep{cwfsr91}.  This level of scattering is comparable
to, though somewhat below, that required for the line of sight to
\psr.  The fitting procedure above produces a clump with
$\log\,\mathrm{SM} = 1.4$, but the scattering parameters of the clump
in the Cordes-Lazio model imply $\log\,\mathrm{SM} = 3.6$.

Although it seems reasonable to attribute some fraction of the
scattering toward \psr\ to \objectname[]{G33.418$-$0.004}, the
\ion{H}{2} region does not appear to be capable of explaining it all.
The line of sight to \psr\ would have to pierce several hundred
similar \ion{H}{2} regions.  Higher density filaments or
tendrils from \ion{H}{2} region may be in front of the pulsar and
responsible for the anomalous scattering, but the identification of
such filaments must await higher resolution and sensitivity H$\alpha$
observations.

Alternately, a larger SM would be obtained if the outer scale in
\objectname[]{G33.418$-$0.004} were $l_0 \ll 10^{-4}$~pc ($l_0 \ll
20$~AU).  Given that the size of \objectname[]{G33.418$-$0.004} is
probably comparable to~1~pc, it is difficult to understand why the
outer scale would be so much smaller.  Moreover, the \ion{H}{2} region
is well offset from the location of \src\ \citep{adps79}, so the
enhanced scattering along the line of sight to the extragalactic
source cannot be attributed to the \ion{H}{2} region.  As
\cite{smbc86} concluded, the enhanced scattering toward \src\ (or
\psr) cannot be attributed to any single object or region.

\section{Conclusions}\label{sec:conclude}

I have observed the extragalactic source \src\ at frequencies
between~0.33 and~15~GHz with the VLA and VLBA and re-analyzed archival
VLA observations between~0.33 and~4.9~GHz.  The source has a complex
structure, as \cite{fsc91} suggested, with components of emission both
to the southwest and northeast of the central component.  However, at
least to~15~GHz, the central component is compact and highly
scattered; our larger frequency range, nearly two orders of magnitude,
extends the earlier analyses of \cite{smbc86} and \cite{fsc91}, who
also found the source to be highly scattered.

I have analyzed the interferometric visibilities to extract parameters
of the spectrum of density fluctuations responsible for scattering and
compared the scattering of \src\ to other nearby sources.  The density
fluctuations are anisotropic, with an anisotropy of about 1.3,
comparable to that seen for other highly scattered sources in the
Galactic disk but less than that toward Galactic center sources.
While the data are suggestive that the position angle of the
anisotropy changes with frequency, the uncertainties are large.
Nonetheless, it seems clear that density fluctuations are becoming
anisotropic on scales of order
$10^{15}\,\mathrm{cm}\,(D/5\,\mathrm{kpc})$.  An inner scale to the density spectrum also appears necessary to
reproduce the observed visibilities.  The evidence for an inner
scale~$l_1$ comes both from direct fits for a density spectrum
including an inner scale as well as from assessing the values of the
density spectral index~$\alpha$ for a power-law--only density
spectrum.  Both of these approaches suggest an inner scale of roughly
a few hundred kilometers, comparable to that found toward other
sources.

The two observations at~0.33~GHz span approximately a decade.  There
is essentially no change in the shape of the image over this interval.
A conservative upper limit on the velocity of the scattering material
is 1800~km~s${}^{-1}$.  This value is sufficiently large that it does
not exclude material from the supernova remnant
\objectname[]{G33.6$+$0.1} from contributing to the enhanced
scattering along this line of sight.

In the 0.33~GHz image of the field around \src, there is a number of
other sources that might also be heavily scattered.  As is seen in
other regions of intense scattering (e.g., the Cygnus region and the
Galactic center), there are large changes in the strength of
scattering on lines of sight separated by a degree or less.  With only
a single frequency to estimate the scattering, not all of the large
diameters in this region can be attributed to scattering.
Nonetheless, many of the sources probably are heavily scattered, and
these other sources in the field could serve as a useful starting
point for future scattering observations in this direction.

Both \src\ and \psr, which are separated by only 13\arcmin, show
extremely strong scattering, and \psr\ shows an anomalous frequency
scaling for its pulse broadening between~1.4 and~2.7~GHz.  While there
is no indication of an anomalous frequency scaling for the angular
broadening of \src\ above~5~GHz, the observations between~1.4
and~2.7~GHz (namely those at~1.5 and~2.3~GHz) are insufficient to
place strong constraints on the presence of any anomalous frequency
scaling.  Moreover, the observations were insufficient for placing
useful constraints on the angular broadening of \psr; simultaneous
observations of \psr\ and \src\ at~1.6~GHz are under way and will be
reported elsewhere.

If the lines of sight to the pulsar and extragalactic source are
affected by the same ``clump'' of scattering material, it must be at
least 2.3~kpc distant.  However, a detailed attempt to account for all
of the scattering observables toward these sources (angular broadening
of the extragalactic source, pulse broadening of the pulsar, and upper
limits on the angular broadening of the pulsar) does not produce a
self-consistent set of parameters for a clump that can reproduce all
three measured scattering observables.

There is a clump of H$\alpha$ emission that lies in between the lines
of sight to \psr\ and \src, which is associated with an unobscured
part of the \ion{H}{2} region \objectname[]{G33.418$-$0.004}.
However, unless conditions within this \ion{H}{2} region are unusual,
it appears unable to produce a sufficient amount of scattering to
explain the excess scattering along both (or even either) lines of
sight.  As \cite{smbc86} concluded, the scattering toward either \src\
or \psr\ cannot be attributed to a single object or region.

\acknowledgements
I thank R.~Mutel for the initial encouragement to undertake this
project, S.~Spangler for helpful discussions regarding scattering and
structure functions and for providing his fitting routines, A.~Fey for
discussions about the structure of \src, the referee for a number of
comments that improved the presentation of this work, and D.~Boboltz,
A.~Fey, and R.~Gaume for their hospitality at the US Naval Observatory
where much of the analysis was performed.  I thank the individuals
involved in the Southern H-Alpha Sky Survey Atlas (SHASSA), which is
supported by the National Science Foundation, for making their results
easily available.  This research has made use of NASA's Astrophysics
Data System Bibliographic Services and of the SIMBAD database,
operated at CDS, Strasbourg, France.  The VLA and the VLBA are
facilities of the National Science Foundation operated under
cooperative agreement by Associated Universities, Inc.  A portion of
this work was conducted while the author was a National Research
Council Research Associate at the \hbox{NRL}.  Basic research in radio
astronomy at the NRL is supported by the Office of Naval Research.


\begin{thebibliography}{}
\bibitem[\protect\citeauthoryear{Alberdi et al.}{1993}]{alberdietal93}
	Alberdi, A., Lara, L., Marcaide, J.~M., Elosegui, P., Shapiro,
	I.~I., Cotton, W.~D., Diamond, P.~J., Romney, J.~D., \&
	Preston, R.~A  1993, \aap, 277, L1

\bibitem[\protect\citeauthoryear{Altenhoff et al.}{1979}]{adps79}
	Altenhoff, W.~J., Downes, D., Pauls, T., \& Schraml, J.  1979, \aaps, 35, 23

\bibitem[\protect\citeauthoryear{Armstrong, Rickett, \&
	Spangler}{Armstrong et al.}{1995}]{ars95} Armstrong, J.~W.,
	Rickett, B.~J., \& Spangler, S.~R.  1995, \apj, 443, 209

\bibitem[\protect\citeauthoryear{Briggs et al.}{Briggs, Schwab, \&
	Sramek}{1999}]{bss99} Briggs, D.~S., Schwab, F.~R., \& Sramek,
	R.~A.  1999, in Synthesis Imaging in Radio Astronomy~II, eds.\
	G.~B.~Taylor, C.~L.~Carilli, \& R.~A.~Perley (ASP: San
	Francisco) p.~127

\bibitem[\protect\citeauthoryear{Chandran \& Backer}{2002}]{cb02}
	Chandran, B.~D.~G.\ \& Backer, D.~C.  2002, \apj, 576, 176

\bibitem[\protect\citeauthoryear{Clegg, Fiedler, \& Cordes}{Clegg et al.}{1993}]{cfc93} Clegg, A.~W., Fiedler, R.~L., \&
        Cordes, J.~M. 1993, \apj, 409, 691

\bibitem[\protect\citeauthoryear{Clifton et al.}{1988}]{cfkw88}
	Clifton, T.~R., Frail, D.~A., Kulkarni, S.~R., \& Weisberg,
	J.~M.  1988, \apj, 333, 332

\bibitem[\protect\citeauthoryear{Coles et al.}{1987}]{cfrc87}
	Coles, W.~A., Rickett, B.~J., Codona, J.~L., \& Frehlich, R.~G.   1987, \apj, 315, 666

\bibitem[\protect\citeauthoryear{Condon et al.}{1998}]{ccgyptb98} Condon, J.~J., Cotton, W.~D.,
	Greisen, E.~W., Yin, Q.~F., Perley, R.~A., Taylor, G.~B., \&
	Broderick, J.~J.  1998, \aj, 115, 1693

\bibitem[\protect\citeauthoryear{Cordes \& Lazio}{2003}]{cl03} Cordes,
	J.~M.\ \& Lazio, T.~J.  2003, \apj, submitted

\bibitem[\protect\citeauthoryear{Cordes \& Lazio}{2001}]{cl01} Cordes,
	J.~M.\ \& Lazio, T.~J.  2001, \apj, 549, 997

\bibitem[\protect\citeauthoryear{Cordes \& Lazio}{1991}]{cl91} Cordes,
	J.~M.\ \& Lazio, T.~J.  1991, \apj, 376, 123

\bibitem[\protect\citeauthoryear{Cordes et al.}{1991}]{cwfsr91}
	Cordes, J.~M., Weisberg, J.~M., Frail, D.~A., Spangler, S.~R.,
	\& Ryan, M.  1991, \nat, 354, 121

\bibitem[\protect\citeauthoryear{Cordes et al.}{1985}]{cwb85} Cordes, J.~M., Weisberg, J.~M., \&
	Boriakoff, V.  1985, \apj, 288, 221

\bibitem[\protect\citeauthoryear{Cornwell \& Perley}{1992}]{cp92} Cornwell, T.~J.\ \& Perley,
	R.~A.  1992, \aap, 261, 353

\bibitem[\protect\citeauthoryear{Dame, Hartmann, \& Thaddeus}{Dame et al.}{2001}]{dht01}
	Dame, T.~M., Hartmann, D., \& Thaddeus, P.  2001, \apj, 547, 792

\bibitem[\protect\citeauthoryear{Dennison et al.}{1984}]{dbtbbc84}
	Dennison, B., Broderick, J.~J., Thomas, M., Booth, R.~S.,
	Brown, R.~L., \& Condon, J.~J.  1984, \aap, 135, 199

\bibitem[\protect\citeauthoryear{Desai \& Fey}{2001}]{df01} 
	 Desai, K.~M.\ \& Fey, A.~L.  2001, \apjs, 133, 395

\bibitem[\protect\citeauthoryear{Desai, Gwinn, \& Diamond}{Desai et al.}{1994}]{dgd94}
	Desai, K.~M., Gwinn, C.~R., \& Diamond, P.~J.  1994, \nat, 372, 754

\bibitem[\protect\citeauthoryear{Dickey et al.}{1983}]{dkhv83} Dickey,
	J.~M., Kulkarni, S.~R., Heiles, C.~E., \& van~Gorkom, J.~H.
	1983, \apjs, 53, 591

\bibitem[\protect\citeauthoryear{Fey, Spangler, \& Cordes}{Fey et
	al.}{1991}]{fsc91} Fey, A.~L., Spangler, S.~R., \& Cordes,
	J.~M.  1991, \apj, 372, 132

\bibitem[\protect\citeauthoryear{Fey, Spangler, \& Mutel}{Fey et
	al.}{1989}]{fsm89} Fey, A.~L., Spangler, S.~R., \& Mutel,
	R.~L.  1989, \apj, 337, 730

\bibitem[\protect\citeauthoryear{Frail et al.}{1994}]{fdcv94} Frail, D.~A., Diamond, P.~J.,
        Cordes, J.~M., \& van~Langevelde, H. J.  1994, \apj, 427, L43

\bibitem[\protect\citeauthoryear{Frail \& Weisberg}{1990}]{fw90}
	Frail, D.~A.\ \& Weisberg, J.~M.  1990, \aj, 100, 743

\bibitem[\protect\citeauthoryear{F\"urst et al.}{1990}]{frrr90} Furst,
	E., Reich, W., Reich, P., \& Reif, K.  1990, \aaps, 85, 805

\bibitem[\protect\citeauthoryear{Gaustad et al.}{2001}]{gmrv01}
	Gaustad, J.~E., McCullough, P.~R., Rosing, W., \&. Van~Buren,
	D.  2001, \pasp, 113, 1326

\bibitem[\protect\citeauthoryear{Gupta, Rickett, \& Lyne}{Gupta et al.}{1988}]{grl88} Gupta, Y., Rickett, B.~J., \& Lyne,
        A. 1988, in Radio Wave Scattering in the Interstellar Medium,
        ed.\ J.~M.\ Cordes, B.~J.\ Rickett, \& D.~C.\ Backer (New
        York: AIP), 140

\bibitem[\protect\citeauthoryear{Hewish, Wolszczan, \& Graham}{Hewish et al.}{1985}]{hwg85} Hewish, A., Wolszczan, A., \&
        Graham, D.~A. 1985, \mnras, 213, 167

\bibitem[\protect\citeauthoryear{Jauncey et al.}{1989}]{jaunceyetal89}
	Jauncey, D.~L., Tzioumis, A.~K., Preston, R.~A., et al.  1989,
	\aj, 98, 44

\bibitem[\protect\citeauthoryear{Lambert \& Rickett}{1999}]{lr99}
	Lambert, H.~C.\ \& Rickett, B.~J.   1999, \apj, 517, 299

\bibitem[\protect\citeauthoryear{Lazio}{1997}]{l97} Lazio, T.~J.~W.
	1997, \hbox{Ph.D.} thesis, Cornell University

\bibitem[\protect\citeauthoryear{Lazio \& Cordes}{1998}]{lc98} Lazio, T.~J.~W.\ \& Cordes,
	J.~M.  1998, \apj, 505, 715

\bibitem[\protect\citeauthoryear{Lazio, Spangler, \& Cordes}{Lazio et
	al.}{1990}]{lsc90}  Lazio, T.~J., Spangler, S.~R., \& Cordes,
	J.~M.   1990, \apj, 363, 515

\bibitem[\protect\citeauthoryear{Lithwick \& Goldreich}{2003}]{lg03}
	Lithwick, Y.\ \& Goldreich, P.  2003, \apj, 582, 1220

\bibitem[\protect\citeauthoryear{Lithwick \& Goldreich}{2001}]{lg01}
	Lithwick, Y.\ \& Goldreich, P.  2001, \apj, 562, 279

\bibitem[\protect\citeauthoryear{Lo et al.}{1985}]{lbekrm85} Lo,
	K.~Y., Backer, D.~C., Ekers, R.~D., Kellermann, K.~I., Reid,
	M., \& Moran, J.~M.  1985, \nat, 315, 124

\bibitem[\protect\citeauthoryear{Lockman}{1989}]{l89} Lockman, F.~J.   1989, \apjs, 71, 469

\bibitem[\protect\citeauthoryear{L\"ohmer et al.}{2001}]{lkmll01}
	L\"ohmer, O., Kramer, M., Mitra, D., Lorimer, D.~R., \& Lyne,
	A.~G.  2001, \apj, 562, L157

\bibitem[\protect\citeauthoryear{Marcaide et al.}{1999}]{malp-td99}
	Marcaide, J.~M., Alberdi, A., Lara, L., Perez-Torres, M.~A.,
	\& Diamond, P.~J.  1999, \aap, 343, 801

\bibitem[\protect\citeauthoryear{Moran et al.}{1990}]{mgrb90}
        Moran, J.~M., Greene, B., Rodr{\'\i}guez, L.~F., \& Backer,
        D.~C.  1990, \apj, 348, 147

\bibitem[\protect\citeauthoryear{Narayan}{1992}]{n92} Narayan, R.
	1992, Philos.\ Trans.\ R.\ Soc.\ London, 341, 151

\bibitem[\protect\citeauthoryear{Molnar et al.}{1995}]{mmrj95} 
	Molnar, L.~A., Mutel, R.~L., Reid, M.~J., \& Johnston, K.~J.
	1995, \apj, 438, 708

\bibitem[\protect\citeauthoryear{Perley}{1989}]{p89} Perley, R.~A.  1989, in Sythesis Imaging in 
	Radio Astronomy, eds.\ R.~A.~Perley, F.~R.~Schwab, \&
	A.~H.~Bridle (San Francisco: ASP) p.~287

\bibitem[\protect\citeauthoryear{Ramachandran et al.}{1997}]{rmdma97}
	Ramachandran, R., Mitra, D., Deshpande, A.~A., McConnell,
	D.~M., \& Ables, J.~G.  1997, \mnras, 290, 260

\bibitem[\protect\citeauthoryear{Reid et al.}{1999}]{rrvt99} Reid,
	M.~J., Readhead, A.~C.~S., Vermeulen, R.~C., \& Treuhaft,
	R.~N.  1999, \apj, 524, 816

\bibitem[\protect\citeauthoryear{Rickett}{1990}]{r90} Rickett, B.~J.  1990, \araa, 28, 561

\bibitem[\protect\citeauthoryear{Romani, Narayan, \& Blandford}{Romani
	et al.}{1986}]{rnb86}  Romani, R.~W., Narayan, R., \&
	Blandford, R.  1986, MNRAS, 220, 19

\bibitem[\protect\citeauthoryear{Simonetti \& Cordes}{1989}]{sc89}
	Simonetti, J.~H.\ \& Cordes, J.~M.  1989, \baas, 21, 1124

\bibitem[\protect\citeauthoryear{Slysh et al.}{2001}]{slyshetal01}
	Slysh, V.~I., et al.  2001, \mnras, 320, 217

\bibitem[\protect\citeauthoryear{Stanimirovic et
	al.}{2003}]{swdddha03}  Stanimirovic, S., Weisberg, J.~M.,
	Dickey, J.~M., de~la~Fuente, A., Devine, K., Hedden, A., \&
	Anderson, S.~B.  2003, \apj, in press

\bibitem[\protect\citeauthoryear{Spangler \& Cordes}{1998}]{sc98}
	Spangler, S.~R.\ \& Cordes, J.~M.  1998, \apj, 505, 766

\bibitem[\protect\citeauthoryear{Spangler \& Cordes}{1988}]{sc88} Spangler, S.~R.\ \& Cordes,
	J.~M.  1988, in Radio Wave Scattering in the Interstellar
	Medium, eds.\ J.~M.\ Cordes, B.~J.\ Rickett, \& D.~C.\ Backer
	(New York: AIP) p.~117

\bibitem[\protect\citeauthoryear{Spangler \& Gwinn}{1990}]{sg90}
	Spangler, S.~R.\ \& Gwinn, C.~R.  1990, \apj, 353, L29

\bibitem[\protect\citeauthoryear{Spangler et al.}{1986}]{smbc86} Spangler, S.~R., Mutel, R.~L.,
	Benson, J., \& Cordes, J.~M.  1986, \apj, 301, 312

\bibitem[\protect\citeauthoryear{Taylor \& Cordes}{1993}]{tc93}
	Taylor, J.~H.\ \& Cordes, J.~M.  1993, {\apj}, 411, 674

\bibitem[\protect\citeauthoryear{Trotter, Moran, \& Rodr{\'\i}guez}{Trotter et al.}{1998}]{tmr98} Trotter, A., Moran, J., \&
	Rodr{\'\i}guez, L.  1998, \apj, 493, 666

\bibitem[\protect\citeauthoryear{van~Gorkom et al.}{1982}]{vgsg82}
	van~Gorkom, J.~H., Goss, W.~M., Seaquist, E.~R., \& Gilmore,
	W.~S.  1982, \mnras, 198, 757

\bibitem[\protect\citeauthoryear{van~Langevelde et al.}{1992}]{vfcd92} van~Langevelde, H.~J.,
        Frail, D.~A., Cordes, J.~M., \& Diamond, P.~J. 1992, \apj, 396, 686

\bibitem[\protect\citeauthoryear{Wilkinson, Narayan, \&
	Spencer}{Wilkinson et al.}{1994}]{wns94} Wilkinson, P.~N.,
	Narayan, R., \& Spencer, R.~E  1994, \mnras, 269, 67

\bibitem[\protect\citeauthoryear{Wolzszcan \& Cordes}{1987}]{wc87} Wolzszcan, A.\ \& Cordes,
        J.~M.  1987, \apj, 320, L35

\bibitem[\protect\citeauthoryear{Yusef-Zadeh et al.}{1994}]{y-zcwmr94} Yusef-Zadeh, F., Cotton,
        W., Wardle, M., Melia, F., \& Roberts, D.~A.  1994, \apj, 434,
        L63
\end{thebibliography}
\end{document}